\documentclass[a4paper, UKenglish, cleveref, autoref, thm-restate, screen]{lipics-v2021}

\pdfoutput=1 
\hideLIPIcs  

\title{From Tarski to Russell : Bridging Explicit and Implicit presentations of Type Theory}

\author{Thierry Coquand}{Chalmers University of Technology -- University of Gothenburg, Sweden}{coquand@chalmers.se}{https://orcid.org/0000-0002-5429-5153}{}

\author{Raphaël Sterbac}{ENS Paris-Saclay -- Université Paris-Saclay, France}{raphael.sterbac@ens-paris-saclay.fr}{https://orcid.org/0009-0001-6779-8323}{}

\authorrunning{T. Coquand and R. Sterbac} 

\Copyright{Thierry Coquand and Raphaël Sterbac} 



\acknowledgements{The authors are grateful to Meven Lennon-Bertrand and Thiago Felicissimo for their useful feedback on the first drafts of this work, and on the mechanized proof.}

\category{} 

\relatedversion{} 

\supplement{Formalisations in Rocq, Agda and Unit \cite{Formalisation,auto-formalisation,unit-formalisation}}

\nolinenumbers 

\EventEditors{John Q. Open and Joan R. Access}
\EventNoEds{2}
\EventLongTitle{42nd Conference on Very Important Topics (CVIT 2016)}
\EventShortTitle{CVIT 2016}
\EventAcronym{CVIT}
\EventYear{2016}
\EventDate{December 24--27, 2016}
\EventLocation{Little Whinging, United Kingdom}
\EventLogo{}
\SeriesVolume{42}
\ArticleNo{23}

\usepackage{ebproof}
\usepackage{ebproof-rules}
\usepackage{mathpartir}
\usepackage{verbatim}
\usepackage{float}
\usepackage{circuitikz}
\usepackage{stmaryrd}

\usepackage[dvipsnames]{xcolor}
\definecolor{Bg}{HTML}{F9F9F9}
\definecolor{link}{HTML}{771e7e}
\usepackage{tcolorbox}
\tcbuselibrary{skins,breakable}
\newtcolorbox{rulebox}[1][]{
  enhanced,
  breakable,
  frame hidden,
  arc=2pt,
  outer arc=2pt,
  colframe=black,
  coltitle=black,
  detach title,
  before upper={\tcbtitle\par\vspace{1ex}},
  top=3pt,
  bottom=0pt,
  left=3pt,
  right=3pt,
  colback=Bg,
  #1
}
\usepackage{hyperref}
\hypersetup{
    colorlinks=true,
    linkcolor=link,
    filecolor=link,      
    urlcolor=link,
    citecolor=link
}

\usepackage{tikz-cd}
\usepackage{amsthm}
\usepackage{amsmath}
\usepackage{amssymb} 

\usepackage{zref-clever}
\zcsetup{cap}

\zcRefTypeSetup{section}{
  Name-sg = \S,
  Name-pl = \S\S,
  name-sg = \S,
  name-pl = \S\S,
}

\zcRefTypeSetup{construction}{
  Name-sg = Construction,
  Name-pl = Constructions,
  name-sg = construction,
  name-pl = constructions
}

\newcommand{\bN}{\mathbb{N}}

\usepackage{xspace}
\definecolor{color-GAT}{HTML}{000000}
\definecolor{color-Tarski}{HTML}{1080bf}
\definecolor{color-Russell}{HTML}{db6840}
\definecolor{color-PTS}{HTML}{397f32}

\newcommand{\ThyGAT}{\textcolor{color-GAT}{\textbf{$\mathcal{T}_\text{GAT}$}}\xspace}
\newcommand{\ThyTarski}{\textcolor{color-Tarski}{\textbf{$\mathcal{T}_\text{Tarski}$}}\xspace}
\newcommand{\ThyRussell}{\textcolor{color-Russell}{\textbf{$\mathcal{T}_\text{Russell}$}}\xspace}
\newcommand{\ThyPTS}{\textcolor{color-PTS}{\textbf{$\mathcal{T}_\text{PTS}$}}\xspace}
\newcommand{\ThyE}{\textbf{$\mathcal{T}_\text{E}$}\xspace}

\newcommand{\vdashG}{\mathrel{\textcolor{color-GAT}{\vdash_{\textbf{\tiny G}}}}}
\newcommand{\vdashT}{\mathrel{\textcolor{color-Tarski}{\vdash_{\textbf{\tiny T}}}}}
\newcommand{\vdashR}{\mathrel{\textcolor{color-Russell}{\vdash_{\textbf{\tiny R}}}}}
\newcommand{\vdashP}{\mathrel{\textcolor{color-PTS}{\vdash_{\textbf{\tiny P}}}}}
\newcommand{\vdashE}{\vdash_{\mathsf{E}}}

\newcommand{\eraseT}[1]{\textcolor{color-Tarski}{\lvert} #1 \textcolor{color-Tarski}{\rvert_{\mathsf{T}}}}
\newcommand{\eraseR}[1]{\textcolor{color-Russell}{\lvert} #1 \textcolor{color-Russell}{\rvert_{\mathsf{R}}}}
\newcommand{\eraseP}[1]{\textcolor{color-PTS}{\lvert} #1 \textcolor{color-PTS}{\rvert_{\mathsf{P}}}}
\NewDocumentCommand{\liftG}{o m}{\lceil #2 \rceil\IfValueT{#1}{_{#1}}}

\newcommand{\coe}{\mathsf{coe}}
\newcommand{\JoinTy}{\mathsf{Join}}
\newcommand{\nf}{\mathsf{nf}}
\newcommand{\ctx}{\mathsf{ctx}}

\newcommand{\Elem}{\mathsf{Tm}}
\newcommand{\Type}{\mathsf{Ty}}

\newcommand{\type}{\mathsf{type}}

\newcommand{\app}{\mathsf{app}}

\newcommand{\Id}{\mathsf{id}}
\newcommand{\p}{\mathsf{p}}
\newcommand{\q}{\mathsf{q}}

\newcommand{\U}{\mathsf{U}}

\usepackage[square,numbers]{natbib}

\ebproofset{
    rule margin=0.5ex,
    separation=1em
}
\NewDocumentEnvironment{mathparsmall}{}
  {%
    \small
    \begin{mathparpagebreakable}
  }
  {%
    \end{mathparpagebreakable}%
  }

\NewDocumentEnvironment{mathpartiny}{}
  {%
    \scriptsize
    \begin{mathparpagebreakable}
  }
  {%
    \end{mathparpagebreakable}%
  }

\DeclareDocumentCommand{\coqlogo}{}{\raisebox{-.2em}{\includegraphics[height=.9em]{icon-rocq-blue-eps-converted-to.pdf}}}
\DeclareDocumentCommand{\formfile}{O{blob/8920605948f1c2d0b92fd81ace628cb6996f46fe/} m}{%
  \href{https://github.com/raphael-sterbac/Russell-Tarski-Equivalence/\IfValueT{#1}{#1}#2}{\coqlogo}}
\DeclareDocumentCommand{\formline}{O{blob/8920605948f1c2d0b92fd81ace628cb6996f46fe/} m m m}{%
  \href{https://github.com/raphael-sterbac/Russell-Tarski-Equivalence/\IfValueT{#1}{#1}#2\#L#3}{\coqlogo}}

\DeclareDocumentCommand{\agdalogo}{}{\raisebox{-.2em}{\includegraphics[height=.9em]{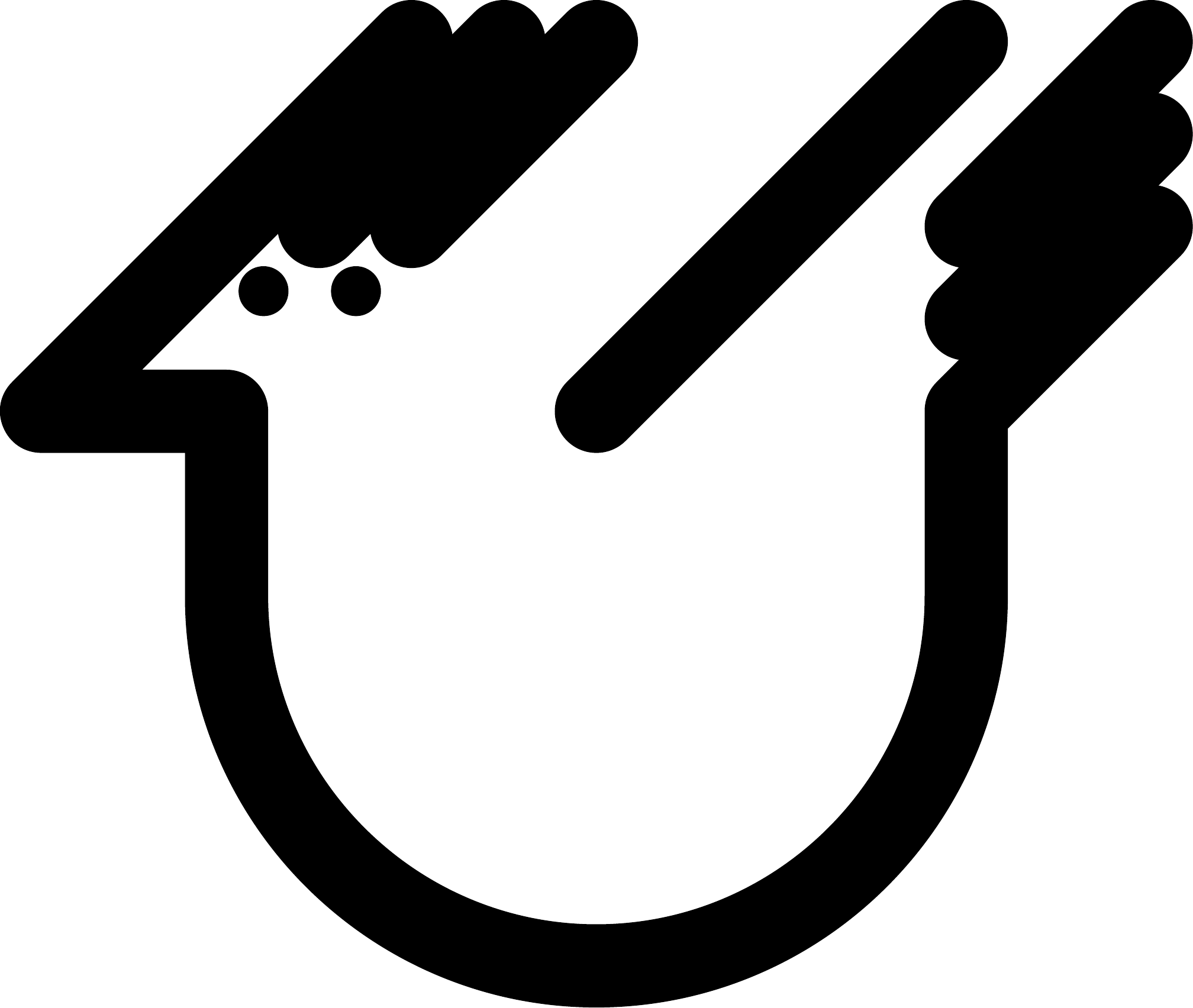}}}
\DeclareDocumentCommand{\agdalink}{m}{\href{#1}{\agdalogo}}
\begin{document}

\maketitle

\begin{abstract}
    We present a modular and elementary proof for the equivalence of different
    presentations of type theory, notably relating Russell and Tarski presentations of cumulative universe hierarchies. More generally, we relate the generalised algebraic and pure type system formulations of the syntax, bridging the gap between the type theory used to prove metatheoretical results, and the one used in practical proof systems. Our work extends initiality results to implicit presentations of type theory.
\end{abstract}

{\small
\par\addvspace{0.5\baselineskip}%
{\noindent \sffamily\bfseries\color{darkgray}Use of Artificial Intelligence} All definitions, proofs, and results of this paper were conceived by the authors, and the paper has been completely human-written. During the research process, we used Claude to develop a formalisation of the paper in Agda \cite{auto-formalisation} and Unit \cite{unit-formalisation, Coquand2026Unit}. This notably helped us detect a mistake in \zcref{sectionT2T2'}, and verify sections \zcref{extension} and \zcref{sec:semantics}. The formalisation was audited by the authors to ensure faithfulness and the absence of postulates.}

\section{Introduction}

In type theory, universes allow manipulating types as terms.
To avoid Girard's paradox, Martin-Löf \cite{Martin-Lof-1972, MARTINLOF75} introduced a \textit{hierarchy} of universes: $\U_0, \U_1, \U_2, ...$.
Those can be assumed to be \textit{cumulative}, requiring proper inclusions between them.
Martin-Löf’s first presentations \cite{MARTINLOF75, MARTINLOF82} were formulated à la Russell, where a universe is seen as a type of types. Later, an alternative version à la Tarski \cite{MARTINLOF84}, where a universe is seen as a type of codes along with a decoding function, was also introduced, and Russell universes can be regarded as an informal notation. 

 When presenting Type Theory in an algebraic fashion, Tarski universes are essential if we do not want to introduce sort equations. Indeed, identifying the sort of terms of a universe with a sort of types forces strict set-theoretic equalities that fail in many semantic models. Connected this issue, Voevodsky proposed the Initiality Conjecture project, which aimed at developing rigorous metatheory for a large class of theories. A first step at solving this goal was achieved by Streicher [29], proving a result which can be interpreted as initiality for a fully annotated Calculus of Constructions, and then proving its equivalence with the unannotated formulation. Generalised Algebraic Theories (GATs) provide a framework to describe type theories and their models in a uniform way. In this paper, we extend Streicher’s equivalence results to relate the Generalised Algebraic and Pure Type System presentations. This provides a rigorous justification of the common practice of using informal unannotated syntax with Russell universes, and legitimates interpreting such syntax in the usual models of type theory.

\noindent
\paragraph*{Related work}
Assaf \cite{assaf} showed the equivalence between Russell and Tarski presentations of universes for the Calculus of Constructions with cumulativity and $\beta$ as an external reduction. A proof extending this result to cumulativity with product covariance was recently published by Felicissimo and Winterhalter \cite{tiago-fscd}, using a technique of \citet{Streicher}. Unfortunately, because these approaches rely on confluence and view $\beta$ as an external reduction, their arguments do not apply to formulations of type theory with judgemental equality, and do not extend to $\eta$-equivalence. This justifies the study of a new proof technique, which applies directly to equational presentations of type theory with $\beta\eta$-conversion.

In the present paper, we will study the simplest form of cumulativity, differing from \citet{tiago-fscd} who retain product covariance. Indeed, structural sub-typing breaks canonicity when adding inductive types \cite{Luo}, and product contravariance does not hold in the set model. Also, it is essential to note that both product covariance and contravariance can be simulated with eta expansion in our setting. Yet, stronger forms of structural sub-typing would provide a more convenient manipulation of universes in practical systems \cite{subtyping}. The recent work of \citet{AdapTT} presents a general framework which would provide automatically the necessary equations to solve the issues mentioned above. 

\paragraph*{Contributions} In this paper we relate the Generalised Algebraic presentation of type theory to the implicit presentation inherited from Pure Type Systems, by a method due to \citet{Streicher}. We considered Martin-Löf's Type Theory with dependent products, cumulative universes and judgmental $\beta\eta$ conversion. Our proof is modular and adapts to various extensions of type theory, such as universe polymorphism \cite{polymorphism}.
Concretely, our equivalence result entails that any semantic model $\mathcal{M}$ of the GAT interprets derivations of the implicit, unannotated Russell theory.
This bridges the gap between the correctness and semantics of explicit presentations, and the implicit theories actually implemented in proof systems.
Additionally, this relates modern semantic results to the literature where implicit syntax was the norm.

The main argument is as follows: there are forgetful maps from the implicit theories to the
implicit presentation, and we show that any implicit judgement can be lifted back in a unique way.
To prove this, we rely on well-known properties of the syntax : no-confusion and injectivity of type formers, which are sometimes referred to as \textit{definitional inversion} \cite{definitional-inv}.
We prove them for the generalised algebraic theory, and they will propagate along our equivalence results, differing from \citet{tiago-fscd} who need injectivity for each theories.
Those properties seem to be strongly related to conversion-checking \cite{meven}, and our equivalence result itself is to be related to the elaboration in proof assistants.

Our proof is elementary and provides an algorithm to translate from one presentation to the other. The definitional inversion properties are proven using the new domain-theoretic technique of \citet{definitional-inv}, making our proof adaptable to various theories which do not enjoy normalisation.
In fact, the proof is purely syntactic and can be done internal to a weak metatheory such as primitive recursive arithmetic, whereas normalisation relies on a stronger metatheory such as the constructive set theory of \citet{Coq19, aczel}.

We provide a human-written formalisation of section \zcref{sectionT1T2} in Rocq \cite{Formalisation}, assuming the definitional inversion properties. In addition, we provide a complete auto-formalisation of the paper in Agda \cite{auto-formalisation}, and of the interpretation section (\zcref{sec:semantics}) in Unit \cite{unit-formalisation, Coquand2026Unit}. Throughout the paper, links to the developments are indicated by the logos \includegraphics[height=.9em]{icon-rocq-blue-eps-converted-to.pdf} and \agdalogo.

\paragraph*{Structure of this paper}
In \zcref{GAT} we present the fully explicit Generalised Algebraic Theory and prove definitional inversion, before presenting the zoo of implicit presentations in \zcref{implicit-presentations}.
Then, we prove that all those syntaxes are equivalent in the following subsections: 
\vspace{0.5em}
\begin{itemize}
    \item[] \textbf{\S\S~\ref{sectionT0T1}.} We relate explicit and implicit substitutions. 
    \item[] \textbf{\S\S~\ref{sectionT1T2}.} We relate Tarski and Russell presentations for cumulative universes.
    \item[] \textbf{\S\S~\ref{sectionT2T2'}.} We relate annotated and unannotated syntaxes. 
\end{itemize}
\vspace{0.5em}
Finally, we extend initiality and give semantics to implicit syntax in \zcref{sec:semantics}, before extending our results to universe polymorphism in \zcref{extension}.

\section{The Generalised Algebraic Theory and Definitional inversion}\label{GAT}

\subsection{A fully explicit presentation}

In this section, we define the fully explicit theory with Tarski-style universes as a generalised algebraic theory \ThyGAT.
This means that we present the syntax in an algebraic way, and this provides a canonical notion of model in categories with families (CwF). Those
were originally described by \citet{Cartmell}, and extended with dependent products and universes by \citet{CWF-CCD}.
The theory described here is really close to the one of \citet{bcde2026}.

\begin{definition}[Sorts]
We have four sorts: contexts, substitutions, types, and elements. 
\begin{align*}
    &\vdashG \ctx\\
    \Delta, \Gamma : \ctx &\vdashG \Delta \rightarrow \Gamma\\
    \Gamma : \ctx &\vdashG \Type(\Gamma)\\
    \Gamma : \ctx, A : \Type(\Gamma) &\vdashG \Elem(\Gamma,A)
\end{align*}
\end{definition}

\begin{definition}[Substitutions]
    \ThyGAT has substitutions with composition, identity and an empty context: 

    \noindent
    \begin{minipage}{0.35\linewidth}
        \begin{align*}
            &\vdashG 1 : \ctx\\
            \Gamma &\vdashG ()_\Gamma : \Gamma \rightarrow 1
        \end{align*}
    \end{minipage}
    \hfill
    \begin{minipage}{0.50\linewidth}
        \begin{align*}
            \Gamma : \ctx &\vdashG \Id_\Gamma : \Gamma \rightarrow \Gamma\\
            \Theta, \Delta, \Gamma : \ctx, \sigma : \Delta \rightarrow \Gamma, \delta : \Theta \rightarrow \Delta &\vdashG \sigma\delta : \Theta \rightarrow \Gamma
        \end{align*}
    \end{minipage}

    \vspace{1em}

    which satisfies the following equations (we omit the contexts for the sake of simplicity) :

    \begin{minipage}{0.40\linewidth}
        \centering
        $\sigma \; \Id_\Delta = \Id_\Gamma \; \sigma = \sigma$
    \end{minipage}
    \hfill
    \begin{minipage}{0.40\linewidth}
        \centering
        $(\theta \sigma)\delta = \theta (\sigma \delta)$
    \end{minipage}

\end{definition}

Substitutions are acting on types and elements, in the sense of the following definition.

\begin{definition}[Action of substitutions on types and terms]
    \ThyGAT has the following operations:
    \begin{align*}
    \Gamma : \ctx, \Delta : \ctx, A : \Type(\Gamma), \sigma : \Delta \rightarrow \Gamma &\vdashG A \sigma : \Type(\Delta)\\
    \Gamma : \ctx, \Delta : \ctx, A : \Type(\Gamma), \sigma : \Delta \rightarrow \Gamma, a: \Elem(\Gamma,A) &\vdashG a \sigma : \Elem(\Delta, A\sigma)
    \end{align*}

    Satisfying the following equations:

    \begin{minipage}{0.20\linewidth}
        $A \; \Id = A$
    \end{minipage}
    \hfill
    \begin{minipage}{0.20\linewidth}
        $(A\sigma)\delta = A(\sigma \delta)$
    \end{minipage}
    \hfill
    \begin{minipage}{0.20\linewidth}
        $a \; \Id = a$
    \end{minipage}
    \hfill
    \begin{minipage}{0.20\linewidth}
        $(a\sigma)\delta = a (\sigma \delta)$
    \end{minipage}

\end{definition}

We also need context extension, in the sense of the following.

\begin{definition}[Projections]
    \ThyGAT has the following operations:
    \begin{align*}
        \Gamma : \ctx, A : \Type(\Gamma) &\vdashG \Gamma.A : \ctx\\
        \Gamma, \Delta : \ctx, A : \Type(\Gamma), \sigma : \Delta \rightarrow \Gamma, a: \Elem(\Delta,A\sigma) &\vdashG (\sigma,a) : \Delta  \rightarrow \Gamma.A\\
        \Gamma : \ctx, A : \Type(\Gamma) &\vdashG \p_{\Gamma,A} : \Gamma.A \rightarrow \Gamma \\
        \Gamma : \ctx, A : \Type(\Gamma) &\vdashG \q_{\Gamma,A} : \Elem(\Gamma.A, A\p_{\Gamma,A})
    \end{align*}

    Satisfying the following equations:
    
    \begin{minipage}{0.20\linewidth}
        $\p(\sigma,a) = \sigma$
    \end{minipage}
    \hfill
    \begin{minipage}{0.20\linewidth}
            $\q(\sigma,a) = a$
    \end{minipage}
    \hfill
    \begin{minipage}{0.20\linewidth}
            $(\sigma,a)\delta = (\sigma \delta, a\delta)$
    \end{minipage}
    \hfill
    \begin{minipage}{0.20\linewidth}
            $(\p,\q) = \Id$
    \end{minipage}

\end{definition}

All of this defines an \textit{explicit} substitution calculus, from which
we can recover the usual presentation of substitution, as defined in the following.

\begin{definition}[Usual syntax for substitutions]
    If $A : \Type(\Gamma)$, $a : \Elem(\Gamma,A)$ and $\sigma : \Delta \rightarrow \Gamma$ we define the following: 
    \begin{gather*}
        [a] = (\Id,a) : \Gamma \rightarrow \Gamma.A\\
        \sigma^+ = (\sigma\p,\q) : \Delta.A\sigma \rightarrow \Gamma.A
    \end{gather*}
\end{definition}

Now, we can add logical structure via type connectives in our theory, starting with dependent products.

\begin{definition}[Dependent product]
    \ThyGAT has the following operations:
     \begin{align*}
         \Gamma : \ctx, A : \Type(\Gamma), B : \Type(\Gamma.A) &\vdashG \Pi(A,B) : \Type(\Gamma)\\
         \Gamma : \ctx, A : \Type(\Gamma), B : \Type(\Gamma.A), b : \Elem(\Gamma.A,B) &\vdashG \lambda(A,B,b) : \Elem(\Gamma,\Pi(A,B))\\
         \Gamma : \ctx, A : \Type(\Gamma), B : \Type(\Gamma.A), f : \Elem(\Gamma, \Pi(A,B)), a : \Elem(\Gamma,A) &\vdashG \app(A,B,f,a) : \Elem(\Gamma,B[a])
     \end{align*}

    Those should be functorial and satisfy the $\beta$ and $\eta$ rules, given by the following equations.
    \noindent
    \begin{minipage}{0.30\linewidth}
        \begin{gather*}
            \Pi(A, B)\sigma = \Pi(A\sigma, B \sigma^+)\\
            \lambda(A,B,b)\sigma = \lambda(A\sigma,B\sigma^+,b\sigma^+)\\
            \app(A,B,\lambda(A,B,b),a) = b[a]
        \end{gather*}
    \end{minipage}
    \hfill
    \begin{minipage}{0.50\linewidth}
        \begin{gather*}
            c = \lambda(A,B, \app(A\p,B\p^+,c\p,\q))\\
            \app(A,B,c,a)\sigma = \app(A\sigma,B\sigma^+,c\sigma,a\sigma)
        \end{gather*}
    \end{minipage}

\end{definition}

We now add cumulative universes to our theory, by requiring a countable hierarchy of universes $\U_0, \U_1,...$
along with explicit codes, lifts and decoding operations. This presentation of universes is referred to as \textit{à la Tarski}.
This fully explicit syntax is the usual way to present type theory with cumulative universes in a (generalised) algebraic way.

\begin{definition}[Tarski universes]
    \ThyGAT has a countable list of universes $\U_n$ with a decoding operation $T_l(\cdot)$, 
    codes for type connectives $\Pi^n$ and $\U_l^m$, and lifting operations $t_l^m$.
    \begin{align*}
            \Gamma : \ctx &\vdashG (\U_l)_\Gamma : \Type(\Gamma)\\
            \Gamma : \ctx &\vdashG (\U_l^m)_\Gamma : \Elem(\Gamma,(\U_m)_\Gamma) \quad (l < m)\\
            \Gamma : \ctx, a : \Elem(\Gamma,(\U_l)_\Gamma) &\vdashG t_l^m(a) : \Elem(\Gamma,(\U_m)_\Gamma) \quad\; (l\leq m)\\
            \Gamma : \ctx, a : \Elem(\Gamma,(\U_l)_\Gamma) &\vdashG T_l(a) : \Type(\Gamma)\\
            \Gamma : \ctx, a : \Elem(\Gamma,(\U_l)_\Gamma),  b : \Elem(\Gamma.T_l(a),(\U_l)_{\Gamma.T_l(a)}) &\vdashG \Pi^l(a,b) : \Elem(\Gamma,(\U_l)_\Gamma)
    \end{align*}
    The decoding and lifting maps must be strict, natural, and should commute with substitution:

    \noindent
    \begin{minipage}{0.45\linewidth}
        \begin{gather*}
            T_l(t_n^l(X)) = T_n(X) \text{ for } n \leq l \\
            T_n(\Pi^n(X,Y)) = \Pi(T_n(X), T_n(Y)) \text{ for } n \geq 0\\
            T_n(\U_m^n) = \U_m \text{ for } m < n\\
            (\U_l)_\Gamma\sigma = (\U_l)_\Delta \\
            (\U_l^m)_\Gamma\sigma = (\U_l^m)_\Delta \text{ for } l < m \\
            t_l^m(a)\sigma = t_l^m(a\sigma) \text{ for } l \leq m
        \end{gather*}
    \end{minipage}
    \hfill
    \begin{minipage}{0.50\linewidth}
        \begin{gather*}
            t_n^n(X) = X \\
            t_n^p(t_m^n(X)) = t_m^p(X) \text{ for } m \leq n \leq p \\
            t_n^p(\Pi^n(X,Y)) = \Pi^p(t_n^p(X),t_n^p(Y)) \text{ for } n \leq p \\
            t_n^p(\U_m^n) = \U_m^p \text{ for } m < n \leq p\\
            T_l(a)\sigma = T_l(a\sigma) \\
            \Pi^l(a,b)\sigma = \Pi^l(a\sigma, b\sigma^+)
        \end{gather*}
    \end{minipage}
\end{definition}

It is important to note that those equations have been carefully chosen so that this presentation is equivalent to the \textit{implicit} Russell universes. In fact, those naturality and functoriality equations are already well-known and already appear in the work of \citet{assaf, thire}, where they are described as requiring \textit{`full reflection''}. This presentation is intrinsically typed \cite{CWF-CCD,QIIT}, but in the following we will work with the corresponding extrinsic system, which is defined as a deductive system over raw terms. 

\subsection{Injectivity and no-confusion}

In the next sections, we will rely on three definitional inversion properties: injectivity of $\Pi$-types, injectivity of universes, and no-confusion between $\Pi$ and $\U_n$. 
Rather than proving them as corollaries of normalisation (which requires a strong metatheory) or via untyped confluence (which fails in the presence of judgemental $\eta$-equality), we establish them using the new domain-theoretic technique of \citet{definitional-inv}. 
We refer the reader to \citet[Section 2]{definitional-inv} for the general construction and focus on the adaptation to a cumulative hierarchy of universes. The details of the proof can be found in the Agda formalisation \cite{auto-formalisation}.

In \citet{definitional-inv}, terms are interpreted in a Scott Domain $D$, and universe hierarchies are modelled  by erasing levels into a single semantic code $\U$ and recovering universe injectivity from the logical relation \cite[Section 3]{definitional-inv}. Here, to take our cumulative hierarchy into account, we equip the domain with a hierarchy of universe codes $\U_n$ ($n \in \bN$).
\begin{itemize}
    \item \textit{Cumulativity and Tarski coercions.} Because semantic membership in $\U_n$ ignores the level $n$, cumulativity holds trivially in the model. Moreover, we interpret decodings $T_n(a)$ and lifts $t_n^p(a)$ as the identity on semantic elements, and codes $\Pi^n(a,b), \U_m^n$ are interpreted by the same semantic constructors as $\Pi(A,B), \U_m$. Thus, all the naturality and functoriality equations hold by definition.
    \item \textit{Universe injectivity and no-confusion.} Because the codes $\U_n$ and $\Pi(a,b)$ are already distinct in the domain, soundness of the model in the constant bottom environment directly gives \zcref{univ-inj} and \zcref{no-confusion}.
\end{itemize}
Finally, to prove $\Pi$-injectivity in open contexts, we use the same proof technique using logical relations as in \citet[Section 2.6]{definitional-inv}. 
This gives the following theorems, that we will use extensively throughout the paper.

\begin{theorem}[\agdalink{https://github.com/coquand/domain-semantics/blob/main/ERT/PiInjectivityR.agda} Injectivity of dependent product]\label{Pi-inj}
    The dependent product is injective in \ThyGAT: if $\Gamma \vdashG \Pi(A,B) = \Pi(A',B')$, then $\Gamma \vdashG A = A'$ and $\Gamma.A \vdashG B = B'$.
\end{theorem}

\begin{theorem}[\agdalink{https://github.com/coquand/domain-semantics/blob/main/ERT/Injectivity.agda} Injectivity of universes]\label{univ-inj}
    If we have $\Gamma \vdashG \U_{n_0} = \U_{n_1}$ in \ThyGAT, then $n_0 = n_1$. 
\end{theorem}

\begin{theorem}[\agdalink{https://github.com/coquand/domain-semantics/blob/main/ERT/Injectivity.agda} No-Confusion]\label{no-confusion}
    For any types $A$ and $B$, and any level $n$, the following judgement is not derivable in \ThyGAT: $\Gamma \vdashG \Pi(A,B) = \U_n$. 
\end{theorem}

\section{A Zoo of Syntaxes}\label{implicit-presentations}

\subsection{Implicit presentations}

In this section, we introduce the more implicit presentations of type theory, which we will prove equivalent to \ThyGAT.
All those systems will have \emph{implicit substitutions}, meaning that it is defined as a meta-operation rather than as an operation of the theory.

More precisely, a substitution is a map $\alpha : \bN \to \mathsf{Tm}$ assigning a raw term to each de Bruijn index, with action $A[\alpha]$ and $a[\alpha]$ defined inductively. We define: 
\begin{align*}
    \mathsf{id}(n) &:= v_n, &
    \mathsf{wk}(n) &:= v_{n+1}, &
    (\alpha \circ \beta)(n) &:= (\alpha(n))[\beta], \\
    (\alpha, u)(0) &:= u, &
    (\alpha, u)(n+1) &:= \alpha(n), &
    \alpha^+ &:= (\alpha \circ \mathsf{wk}, v_0), \\
    e\p &:= e[\mathsf{wk}], &
    e\p^+ &:= e[\mathsf{wk}^+], &
    e[a] &:= e[(\mathsf{id}, a)].
\end{align*}
Typing $\Delta \vdash \alpha : \Gamma$ and conversion $\Delta \vdash \alpha = \alpha' : \Gamma$ are defined by:

\begin{mathparsmall}
  \ebrule{
    \hypo{\Delta \vdash}
    \infer1{\Delta \vdash \alpha : 1}
  }
  \and
  \ebrule{
    \hypo{\Gamma \vdash A \;\type}
    \hypo{\Delta \vdash \mathsf{wk} \circ \alpha : \Gamma}
    \hypo{\Delta \vdash \alpha(0) : A[\mathsf{wk} \circ \alpha]}
    \infer3{\Delta \vdash \alpha : \Gamma.A}
  }
  \and
  \ebrule{
    \hypo{\Delta \vdash}
    \infer1{\Delta \vdash \alpha = \alpha' : 1}
  }
  \and
  \ebrule{
    \hypo{\Gamma \vdash A \;\type}
    \hypo{\Delta \vdash \mathsf{wk} \circ \alpha = \mathsf{wk} \circ \alpha' : \Gamma}
    \hypo{\Delta \vdash \alpha(0) = \alpha'(0) : A[\mathsf{wk} \circ \alpha]}
    \infer3{\Delta \vdash \alpha = \alpha' : \Gamma.A}
  }
\end{mathparsmall}

The theory \ThyTarski is the same deductive system as \ThyGAT, but without context annotations and with implicit substitutions.
A main goal of this paper is to relate \ThyTarski to the implicit formulation \textit{à la Russell} of cumulative universes denoted with \ThyRussell.
The latter conflates types and terms, with implicit inclusion between universes. 
This system cannot be formulated algebraically, as this would involve adding a sort equation and a stratification of the sort of types $\Type_n$ which we would need to be included in each other.
Both \ThyTarski and \ThyRussell have dependent products, lambda abstraction and application, described with the following rules.

\begin{figure}[H]
    \centering
    \setlength{\fboxsep}{10pt}
    \fbox{%
        \begin{minipage}{0.95\textwidth}
\begin{mathparsmall}
  \ebrule{
    \hypo{\Gamma \vdash A \;\type}
    \infer1{\Gamma.A \vdash v_0 : A\p}
  }
  \and
  \ebrule{
    \hypo{\Gamma \vdash v_n:A}
    \hypo{\Gamma \vdash B \;\type}
    \infer2{\Gamma.B \vdash v_{n+1} : A\p}
  }
  \and
  \ebrule{
    \hypo{\Gamma \vdash A \;\type}
    \hypo{\Gamma.A \vdash B \;\type}
    \infer2{\Gamma \vdash \Pi(A, B) \;\type}
  }
  \and
  \ebrule{
    \hypo{\Gamma.A \vdash b: B}
    \hypo{\Gamma.A \vdash B \;\type}
    \hypo{\Gamma \vdash A \;\type}
    \infer3{\Gamma \vdash \lambda(A,B,b) : \Pi (A, B)}
  }
  \and
  \ebrule{
    \hypo{\Gamma \vdash c:\Pi(A, B)}
    \hypo{\Gamma \vdash a: A}
    \hypo{\Gamma.A \vdash B \;\type}
    \hypo{\Gamma \vdash A \;\type}
    \infer4{\Gamma \vdash \app(A,B,c,a) : B[a]}
  }
  \and
  \ebrule{
    \hypo{\Gamma.A \vdash b: B}
    \hypo{\Gamma \vdash a:A}
    \infer2[($\beta$-rule)]{\Gamma \vdash \app(A,B,\lambda(A,B,b),a) = b[a] : B[a]}
  }
  \and
  \ebrule{
    \hypo{\Gamma \vdash f: \Pi(A,B)}
    \infer1[($\eta$-rule)]{\Gamma \vdash f = \lambda(A,B,\app(A\p,B\p^+,f\p, v_0)) : \Pi(A, B)}
  }
\end{mathparsmall}
\end{minipage}%
    }
    \caption{Logical structure for \ThyTarski and \ThyRussell}
    \label{fig:rulestyping}
\end{figure}

Then, the operations of \ThyGAT can be translated as formation and introduction rules for Tarski-style universes and the theory
\ThyTarski. Namely, we need codes for products and universes, lifts and decoding. We describe the rules in the following:

\begin{figure}[H]
    \centering
    \setlength{\fboxsep}{10pt}
    \fbox{%
        \begin{minipage}{0.95\textwidth}
\begin{mathparsmall}
  \ebrule{
    \hypo{\Gamma \vdashT}
    \infer1{\Gamma \vdashT \U_n \;\type}
  }
  \and
  \ebrule{
    \hypo{\Gamma \vdashT a: \U_n}
    \hypo{\Gamma.T_n(a) \vdashT b: \U_n}
    \hypo{\Gamma \vdashT T_n(a) \;\type}
    \infer3{\Gamma \vdashT \Pi^n(a,b): \U_n}
  }
  \and
  \ebrule{
    \hypo{\Gamma \vdashT a:\U_n}
    \infer1{\Gamma \vdashT T_n(a) \;\type}
  }
  \and
  \ebrule{
    \hypo{\Gamma \vdashT}
    \infer1[($m<n$)]{\Gamma \vdashT \U_m^n: \U_n}
  }
  \and
  \ebrule{
    \hypo{\Gamma \vdashT a: \U_m}
    \infer1[($m\leq n$)]{\Gamma \vdashT t_m^n(a): \U_n}
  }
\end{mathparsmall}
\end{minipage}%
    }
    \caption{Universe rules of \ThyTarski}
    \label{fig:tarski-univ}
\end{figure}

The equations of \ThyGAT can be translated as conversion rules for \ThyTarski (\zcref{fig:tarski-conv}). Namely,
we require the decoding and lifting operation to be strict, and natural.

\begin{figure}[H]
    \centering
    \setlength{\fboxsep}{10pt}
    \fbox{%
        \begin{minipage}{0.95\textwidth}
\begin{mathparsmall}
  \ebrule{
    \hypo{\Gamma \vdashT a: \U_n}
    \hypo{\Gamma.T_n(a) \vdashT b: \U_n}
    \infer2{\Gamma \vdashT T_n(\Pi^n(a,b)) = \Pi(T_n(a),T_n(b))}
  }
  \and
  \ebrule{
    \hypo{\Gamma \vdashT a: \U_n}
    \hypo{\Gamma.T_n(a) \vdashT b: \U_n}
    \infer2[($n\leq p$)]{\Gamma \vdashT t_n^p(\Pi^n(a,b)) = \Pi^p(t_n^p(a),t_n^p(b)) : \U_p}
  }
  \and
  \ebrule{
    \hypo{\Gamma \vdashT}
    \infer1[($m<n$)]{\Gamma \vdashT T_n(\U_m^n) = \U_m}
  }
  \and
  \ebrule{
    \hypo{\Gamma \vdashT}
    \infer1[($m<n\leq p$)]{\Gamma \vdashT t_n^p(\U_m^n) = \U_m^p : \U_p}
  }
  \and
  \ebrule{
    \hypo{\Gamma \vdashT a : \U_m}
    \infer1[($m\leq n$)]{\Gamma \vdashT T_n(t_m^n(a)) = T_m(a)}
  }
  \and
  \ebrule{
    \hypo{\Gamma \vdashT a : \U_m}
    \infer1[($m\leq n\leq p$)]{\Gamma \vdashT t_n^p(t_m^n(a)) = t_m^p(a): \U_p}
  }
  \and
  \ebrule{
    \hypo{\Gamma \vdashT a : \U_n}
    \infer1{\Gamma \vdashT t_n^n(a) = a : \U_n}
  }
\end{mathparsmall}
\end{minipage}%
    }
    \caption{Conversion rules of \ThyTarski}
    \label{fig:tarski-conv}
\end{figure}

Now, for the theory \ThyRussell (Russell universes), types and terms are conflated, and thus
we should be able to deduce $\Gamma \vdashR A \; \type$ from $\Gamma \vdashR A : \U_n$. Conversely,
the intuition is that whenever $\Gamma \vdashR A \; \type$ then we should have  : \[\exists n \in \bN, \ \Gamma \vdashR A : \U_n\]

Instead of having codes for each of the type formers, we now close $\U_n$ under the type connectives. Note that, as we
consider \textit{cumulative} universes, we do not take the suprema of levels for the dependent product. Cumulativity itself
is expressed by asking for the proper inclusion of universes : if $\Gamma \vdashR A : \U_n$ then $\Gamma \vdashR A : \U_p$ for $p > n$. 

\begin{figure}[H]
    \centering
    \setlength{\fboxsep}{10pt}
    \fbox{%
        \begin{minipage}{0.95\textwidth}
\begin{mathparsmall}
\ebrule{
    \hypo{\Gamma \vdashR}
    \infer1{\Gamma \vdashR \U_n \;\type}
  }
  \and
  \ebrule{
    \hypo{\Gamma \vdashR A:\U_n}
    \infer1{\Gamma \vdashR A \;\type}
  }
  \and
  \ebrule{
    \hypo{\Gamma \vdashR A: \U_n}
    \hypo{\Gamma.A \vdashR B: \U_n}
    \infer2{\Gamma \vdashR \Pi(A, B): \U_n}
  }
  \and
  \ebrule{
    \hypo{\Gamma \vdashR}
    \infer1[($m<n$)]{\Gamma \vdashR \U_m: \U_n}
  }
  \and
  \ebrule{
    \hypo{\Gamma \vdashR A: \U_n}
    \infer1[($n<p$)]{\Gamma \vdashR A: \U_p}
  }
\end{mathparsmall}
\end{minipage}%
    }
    \caption{Universe rules of \ThyRussell}
    \label{fig:russell-univ}
\end{figure}

Finally, we also consider a system \ThyPTS, which removes redundant type annotations in functional application
and lambda abstraction from \ThyRussell. Thus, $\app(A,B,c,a)$ and $\lambda(A,B,b)$ become $c \; a$ and $\lambda(A,b)$.
This system is closer to the practical syntax of type theory, inherited from the lambda calculus,
and often referred to as the "pure type system" presentation (PTS).
For a full description of the systems, please refer to \zcref{appendix:rules}.

\subsection{Forgetful maps}

Now, we relate the previously defined theories, by exhibiting forgetful translations from the more explicit presentations to the less explicit ones. All those maps are defined over raw untyped terms, and we will prove that they preserve derivations.

First, we define a forgetful map $\eraseT{\cdot}$ from \ThyGAT{} to \ThyTarski{} which makes substitutions implicit by interpreting explicit substitutions of \ThyGAT{} as meta-level substitutions.

\begin{definition}
    We define $\eraseT{\cdot}$ by structural recursion on the raw syntax of \ThyGAT{} (acting homomorphically on $\Pi, \lambda, \app, T_n, t_m^n, \Pi^n$):

    \noindent
    \begin{minipage}{0.42\linewidth}
    \begin{align*}
        \eraseT{(\U_n)_\Gamma} &= \U_n\\
        \eraseT{(\U_m^n)_\Gamma} &= \U_m^n\\
        \eraseT{\q_{\Gamma,A}} &= v_0\\
        \eraseT{\Id_\Gamma} &= \mathsf{id}\\
        \eraseT{()_\Gamma} &= \mathsf{id}
    \end{align*}
    \end{minipage}
    \hfill
    \begin{minipage}{0.48\linewidth}
    \begin{align*}
        \eraseT{A\sigma} &= \eraseT{A}[\eraseT{\sigma}]\\
        \eraseT{a\sigma} &= \eraseT{a}[\eraseT{\sigma}]\\
        \eraseT{\p_{\Gamma,A}} &= \mathsf{wk}\\
        \eraseT{(\sigma, a)} &= (\eraseT{\sigma}, \eraseT{a})\\
        \eraseT{\sigma\delta} &= \eraseT{\sigma} \circ \eraseT{\delta}
    \end{align*}
    \end{minipage}
\end{definition}

Then, we define the next forgetful map, which erases the lift and decoding operation of Tarski-style universes, and sends codes to the corresponding type formers. Concretely, this operation forgets about all the explicit coercions given by the presentation à la Tarski, and leaves everything implicit.

\begin{definition}[\formline{Erasure.v}{10}{Erasurefunction} Universe erasure]
    We define $\eraseR{\cdot}$ inductively by:

    \noindent
    \begin{minipage}{0.40\linewidth}
    \begin{align*}
        \eraseR{\U_n} &= \U_n\\
        \eraseR{\U_m^n} &= \U_m\\
        \eraseR{\app(A,B,c,a)} &= \app(\eraseR{A},\eraseR{B},\eraseR{c},\eraseR{a})\\
        \eraseR{\lambda(A,B,b)} &= \lambda(\eraseR{A},\eraseR{B},\eraseR{b})\\
        \eraseR{v_i} &= v_i
    \end{align*}
    \end{minipage}
    \hfill
    \begin{minipage}{0.40\linewidth}
    \begin{align*}
        \eraseR{T_n(a)} &= \eraseR{a}\\
        \eraseR{t_m^n(a)} &= \eraseR{a}\\
        \eraseR{\Pi^n(a,b)} &= \Pi(\eraseR{a}, \eraseR{b})\\
        \eraseR{\Pi(A, B)} &= \Pi(\eraseR{A}, \eraseR{B})
    \end{align*}
    \end{minipage}
\end{definition}

Finally, we define the map $\eraseP{\cdot}$ which erases type annotations from a term of the system \ThyRussell, and gives a term of \ThyPTS.
This is analogous to the one defined by \citet{Streicher}.

\begin{definition}
    We define $\eraseP{\cdot}$ inductively by :
    \begin{align*}
        \eraseP{\lambda (A,B,b)} &= \lambda(\eraseP{A}, \eraseP{b})\\ 
        \eraseP{\app (A,B,c,a)} &= \eraseP{c} \; \eraseP{a}\\
        \eraseP{v_i} &= v_i \\
        \eraseP{\Pi(A,B)} &= \Pi(\eraseP{A}, \eraseP{B})\\
        \eraseP{\U_n} &= \U_n
    \end{align*}
\end{definition}

In the following, we will prove lemmas of the following shape, following \citet{Streicher}:\\

\textbf{Lemma.} If $\Gamma \vdash t,t' : A$ in $\mathcal{T}$ with $|t| = |t'|$ in $\mathcal{T}'$, then we have : $\Gamma \vdash t = t' : A$ in $\mathcal{T}$.\\

\begin{remark}As mentioned in the introduction, this relates to elaboration in proof assistants. 
Concretely, this lemma means that if two terms are syntactically the same in the implicit user syntax, then they should have the same
computational behaviour in the core system, and thus should be convertible. This is to be compared with a simulation
result, and is often referred to as a \emph{coherence} lemma.
\end{remark}

\begin{figure}[!ht]
\centering
\resizebox{\textwidth}{!}{%
\begin{circuitikz}
\tikzstyle{every node}=[font=\LARGE]

\draw (0,18.75)    rectangle node {\LARGE \ThyGAT}     (3.5,16.25);
\draw (8.5,18.75)  rectangle node {\LARGE \ThyTarski}  (12,16.25);
\draw (17,18.75)   rectangle node {\LARGE \ThyRussell} (20.5,16.25);
\draw (25.5,18.75) rectangle node {\LARGE \ThyPTS}     (29,16.25);

\draw [->, >=Stealth] (3.5,18.25)  -- (8.5,18.25)  node[midway, fill=white]{$\eraseT{\cdot}$};
\draw [->, >=Stealth] (12,18.25)   -- (17,18.25)   node[midway, fill=white]{$\eraseR{\cdot}$};
\draw [->, >=Stealth] (20.5,18.25) -- (25.5,18.25) node[midway, fill=white]{$\eraseP{\cdot}$};

\draw [->, >=Stealth, dashed] (8.5,16.75)  -- (3.5,16.75);
\draw [->, >=Stealth, dashed] (17,16.75)   -- (12,16.75)   node[midway, below=2pt]{$\U$-inj, $\U \neq \Pi$};
\draw [->, >=Stealth, dashed] (25.5,16.75) -- (20.5,16.75) node[midway, below=2pt]{$\U$/$\Pi$-inj, $\U \neq \Pi$};
\end{circuitikz}
}%

\caption{The zoo of syntaxes}
\label{hierarchy-zoo}
\end{figure}
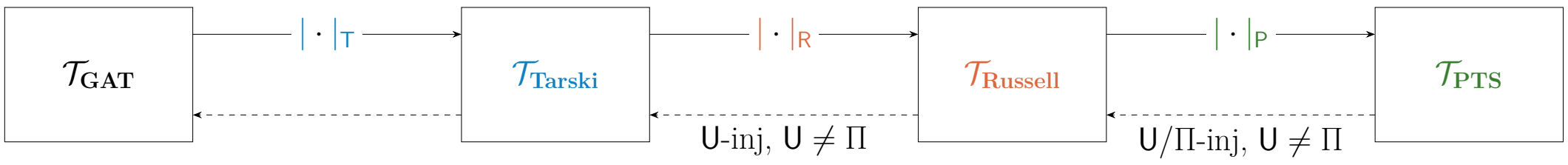

In the next sections, we will prove such lemmas for each of our theories. With those, we will be able to lift a derivation from an implicit theory to an explicit one.
The situation is summarized in \zcref{hierarchy-zoo}, where the strong arrows are the known forgetful maps, and the dotted arrows are the sections we will be constructing, with the necessary definitional inversion properties.
Semantically, this means that their corresponding Categories with Families (CwFs) are all isomorphic to the initial CwF of \ThyGAT.
In particular, any model $\mathcal{M}$ of the GAT in Section 2.1 interprets well-typed derivations of the implicit, unannotated system \ThyPTS. However, those theories crucially do not have the same categories of models, as the decoding operation is not necessarily injective in a model of \ThyGAT. 

As mentioned in the introduction, to prove our equivalence result, we rely on the definitional inversion properties for \textit{only} the explicit system
\ThyGAT proved in \zcref{GAT}, and we propagate those along our equivalence results.
In the following Sections \ref{sectionT0T1}, \ref{sectionT1T2}, and \ref{sectionT2T2'} we show that all the systems \ThyGAT, \ThyTarski, \ThyRussell and \ThyPTS are equivalent.

\section{A Zoo of Equivalences}
\subsection{Explicit and implicit substitutions (\ThyGAT and \ThyTarski)}\label{sectionT0T1}

We now show that \ThyGAT{} and \ThyTarski{} are equivalent, relating the explicit substitution calculus to the presentation where substitution is a meta-operation.

\begin{lemma}[\formline{Utils.v}{710}{substitutionlemma} Substitution lemma]\label{Substitution}
    If $\Gamma \vdashT J$ and $\Delta \vdashT \alpha : \Gamma$, then $\Delta \vdashT J[\alpha]$. Moreover, if $\Delta \vdashT \alpha = \alpha' : \Gamma$, then $\Gamma \vdashT A\;\type$ implies $\Delta \vdashT A[\alpha] = A[\alpha']$ and $\Gamma \vdashT a : A$ implies $\Delta \vdashT a[\alpha] = a[\alpha'] : A[\alpha]$.
\end{lemma}

\begin{theorem}[Soundness of $\eraseT{\cdot}$]
    If $\Gamma \vdashG J$ in \ThyGAT{}, then $\eraseT{\Gamma} \vdashT \eraseT{J}$ in \ThyTarski{}.
\end{theorem}
\begin{proof}
    By mutual induction on derivations in \ThyGAT{}, we check that the erasure of each rule is admissible in \ThyTarski{}. The rules for substitution action and congruence follow immediately from \zcref{Substitution}. Every substitution equation of \ThyGAT{} holds after erasure.

    For instance, the equation $\Delta \vdashG \q_{\Gamma,A}(\sigma, a) = a : A\sigma$ erases to $\eraseT{\q_{\Gamma,A}(\sigma, a)} = v_0[(\eraseT{\sigma}, \eraseT{a})] = \eraseT{a}$, so it holds by reflexivity.
\end{proof}

To lift derivations back to \ThyGAT{} and recover the context annotations, we define a lifting operation $\liftG{-}$, parameterized by the current context $\Gamma$ of \ThyGAT:
\begin{align*}
    \liftG{1} &:= 1 & \liftG{\Gamma.A} &:= \Gamma'.\liftG[\Gamma']{A} \quad (\text{where } \Gamma' := \liftG{\Gamma}) \\
    \liftG[\Gamma.A]{v_0} &:= \q_{\Gamma,A} & \liftG[\Gamma.B]{v_{n+1}} &:= \liftG[\Gamma]{v_n}\p_{\Gamma,B} \\
    \liftG[\Gamma]{\U_n} &:= (\U_n)_\Gamma & \liftG[\Gamma]{\U_m^n} &:= (\U_m^n)_\Gamma\\
    \liftG[\Delta \to 1]{\alpha} &:= ()_\Delta & \liftG[\Delta \to \Gamma.A]{\alpha} &:= (\liftG[\Delta \to \Gamma]{\mathsf{wk} \circ \alpha}, \liftG[\Delta]{\alpha(0)})
\end{align*}

\begin{lemma}[Right inverse]\label{embedding-rightinverse}
    The embedding $\liftG{-}$ is a strict right inverse of $\eraseT{\cdot}$:
    \begin{itemize}
        \item For every context $\Gamma$ of \ThyTarski{}, $\eraseT{\liftG{\Gamma}} = \Gamma$.
        \item For any raw type $A$ in $\Gamma$ of \ThyTarski{}, we have $\eraseT{\liftG[\Delta]{A}} = A$ (where $\Delta := \liftG{\Gamma}$).
        \item  For any raw term $a$ in $\Gamma$ of \ThyTarski{}, we have $\eraseT{\liftG[\Delta]{a}} = a$ (where $\Delta := \liftG{\Gamma}$).
    \end{itemize}
\end{lemma}
\begin{proof}
    By induction, using the equality $\eraseT{\liftG[\Gamma]{v_n}\p_{\Gamma,B}} = v_n[\mathsf{wk}] = v_{n+1}$.
\end{proof}

\begin{lemma}[Correctness of $\liftG{-}$]\label{embedding-correctness}
    The embedding $\liftG{-}$ preserves derivations:
    \begin{itemize}
        \item If $\Gamma \vdashT$, then $\liftG{\Gamma} \vdashG$. More generally, if $\Gamma \vdashT J$ and $\vdashG \Gamma' = \liftG{\Gamma}$, then $\Gamma' \vdashG \liftG[\Gamma']{J}$ (with $\Gamma' \vdashG \liftG[\Gamma']{A} = \liftG[\liftG{\Gamma}]{A}$ for types, and similarly for terms).
        \item If $\Delta \vdashT \alpha : \Gamma$ with $\vdashG \Gamma' = \liftG{\Gamma}$ and $\vdashG \Delta' = \liftG{\Delta}$, then $\vdashG \liftG[\Delta' \to \Gamma']{\alpha} : \Delta' \to \Gamma'$.
        \item For every $\Gamma \vdashT A\;\type$, we have $\Delta' \vdashG \liftG[\Gamma']{A}\,\liftG[\Delta' \to \Gamma']{\alpha} = \liftG[\Delta']{A[\alpha]}$.
        \item For every $\Gamma \vdashT a : A$, we have $\Delta' \vdashG \liftG[\Gamma']{a}\,\liftG[\Delta' \to \Gamma']{\alpha} = \liftG[\Delta']{a[\alpha]} : \liftG[\Delta']{A[\alpha]}$.
    \end{itemize}
\end{lemma}
\begin{proof}
    We first prove the canonical case ($\Gamma' := \liftG{\Gamma}$, $\Delta' := \liftG{\Delta}$) by mutual induction, showing that $\liftG{-}$ preserves derivations and commutes with substitution.

    For $v_n$, we prove commutation by induction on $n$. 
    For binders, it follows from the functoriality equations of \ThyGAT{} and the identity $(\liftG[\Delta' \to \Gamma']{\alpha})^+ = \liftG[\Delta'.A' \to \Gamma'.\tilde{A}]{\alpha^+}$ where $\tilde{A} := \liftG[\Gamma']{A}$ and $A' := \liftG[\Delta']{A[\alpha]}$. Using this commutation property, the rules of \ThyTarski{} involving the meta-substitution translate directly to derivable rules in \ThyGAT{}.
    The extension to arbitrary convertible contexts $\Gamma', \Delta'$ then follows by congruence and context conversion.
\end{proof}

\begin{lemma}[Left inverse modulo conversion]\label{embedding-leftinverse}
    In \ThyGAT, we have:
    \begin{itemize}
        \item $\Gamma \vdashG$ implies $\vdashG \Gamma = \liftG{\eraseT{\Gamma}}$
        \item $\Gamma \vdashG A\;\type$ implies $\Gamma \vdashG A = \liftG[\Gamma]{\eraseT{A}}$
        \item $\Gamma \vdashG a : A$ implies $\Gamma \vdashG a = \liftG[\Gamma]{\eraseT{a}} : A$
        \item $\vdashG \sigma : \Delta \to \Gamma$ implies $\vdashG \sigma = \liftG[\Delta \to \Gamma]{\eraseT{\sigma}} : \Delta \to \Gamma$.
    \end{itemize}
\end{lemma}
\begin{proof}
    By mutual induction. The cases $\Id_\Gamma$ and $\p_{\Gamma,A}$ use $(\p,\q) = \Id$ and $()_1 = \Id_1$. Composition $\sigma\delta$ uses $(\tau,a)\delta = (\tau\delta, a\delta)$ and \zcref{embedding-correctness}.The substitution actions $A\sigma$ and $a\sigma$ follow by congruence and \zcref{embedding-correctness}. The remaining constructors are direct by congruence.
\end{proof}

\begin{theorem}[Section of $\eraseT{\cdot}$]
    If $\Gamma \vdashT J$ in \ThyTarski{}, then there exists a unique up to conversion judgment $\Gamma' \vdashG J'$ in \ThyGAT{} such that $\eraseT{\Gamma'} = \Gamma$ and $\eraseT{J'} = J$.
\end{theorem}
\begin{proof}
    Existence is given by $\Gamma' \vdashG \liftG[\Gamma']{J}$ with $\Gamma' := \liftG{\Gamma}$, using \zcref{embedding-rightinverse} and \zcref{embedding-correctness}. Uniqueness up to conversion follows from \zcref{embedding-leftinverse}.
\end{proof}

\subsection{Tarski and Russell universes (\ThyTarski and \ThyRussell)}\label{sectionT1T2}

We now prove the main result of the paper : equivalence of Tarski-style and Russell-style universes i.e. of \ThyTarski and \ThyRussell.
All the results of this section have been completely formalised \textit{by hand} in Rocq.

First, we show that the erasure operation $\eraseR{\cdot}$ preserves derivations.

\begin{lemma}[\agdalink{https://github.com/coquand/domain-semantics/blob/main/ERT/EraseDeriv.agda} \formline{Erasure.v}{143}{Erasurecorrection} Correctness of erasure]
    If $\Gamma \vdashT J$ in \ThyTarski, then $\eraseR{\Gamma} \vdashR \eraseR{J}$ in \ThyRussell.
\end{lemma}

\begin{proof}
    By mutual induction, we prove the statement for our five different kinds of judgements: well-formation of contexts and types,
    typing, and conversion between types and terms. The proof is mostly direct, and amounts to showing that the
    erasure of each rule of \ThyTarski corresponds to an admissible rule of \ThyRussell.
\end{proof}

Then, we want to show that this forgetful map has an inverse, hence proving the equivalence of the two systems. For this, we will need the usual lemmas, which we all prove simultaneously by a mutual induction on the derivation.

\begin{lemma}[\formline{Utils.v}{330}{Hypothesis} Hypothesis inversion]\label{hyp-inv}
    If we have a derivation $\Gamma. A \vdashT J$, where $J$ is an arbitrary judgment,
    then we necessarily have a derivation $\Gamma \vdashT A \; \type$.
\end{lemma}

\begin{lemma}[\formline{Utils.v}{1050}{Hypothesis} Typing and conversion inversion]

    If we have a derivation $\Gamma \vdashT a : A$ or $\Gamma \vdashT a = b :A$ then we necessarily have a derivation $\Gamma \vdashT A \; \type$.
\end{lemma}

Those two lemmas allow us to reason by inversion on given judgements, as Rocq's inversion
tactic would be stuck with rules such as :
\begin{mathparsmall}
  \ebrule{
    \hypo{\Gamma \vdashT a : A}
    \hypo{\Gamma \vdashT A = B}
    \infer2{\Gamma \vdashT a : B}
  }
\end{mathparsmall}

We also need the following lemma, which will allow us to modify derivations given a conversion judgement.

\begin{lemma}[\formline{Utils.v}{1059}{Hypothesis}]\label{Transform}
    If we have a derivation $\Gamma. A \vdashT J$ and a convertibility derivation $\Gamma \vdashT A = B$,
    then we can construct a derivation $\Gamma. B \vdashT J$.
\end{lemma}

Now, we would want to state the main result, by adapting the arguments of \citet{Streicher}. One could try to prove the
following statement:

\vspace{0.5em}
\textbf{Lemma.} If we have $\Gamma \vdashT u_0 : A_0$ and $\Gamma \vdashT u_1 : A_1$ such that $\eraseR{u_0} = \eraseR{u_1}$
in \ThyRussell, then necessarily they are convertible i.e. $\Gamma \vdashT A_0 = A_1$ and $\Gamma \vdashT u_0 = u_1 : A_0$
\vspace{0.5em}

However, this fails when $u_0$ and $u_1$ are codes in different universes. For instance, we have $\Gamma \vdashT \U_m^n : \U_n$ and $\Gamma \vdashT \U_m^l : \U_l$
which have the same erasure $\eraseR{\U_m^n} = \eraseR{\U_m^l} = \U_m$, but are not convertible (at least if $n \neq l$). Therefore,
we need to prove a more general statement that properly deals with the case when $A_0$ and $A_1$ are universes. In this case,
the two codes should become convertible once lifted to the join of their universes $\U_k$, where $k := \max(n_0, n_1)$. 
Even with this generalised lemma, we would get stuck in the proof for the case of the decoding. Indeed, if we have a code $a: \U_n$ and a type $A$ 
whose erasure is $\eraseR{A} = \eraseR{a}$, then we cannot deduce much on $A$ itself as it could well be any type.
Nevertheless, we still know that $\Gamma \vdashR A : \U_n$ holds in Russell-style because it is the same syntactic term as $a$. This should indicate that $A$ is a small type, i.e. a decoded type, and
in this setting we will prove that $\Gamma \vdashT A = T_n(a)$.
Thus, we shall prove the following lemma. 

\begin{lemma}[\agdalink{https://github.com/coquand/domain-semantics/blob/main/ERT/UniquenessTermPartial.agda} \formline{Equivalence.v}{813}{eraseinjty}] \label{mainuniverse}
    In \ThyTarski, we have:
    \begin{enumerate}
    \item If $\Gamma \vdashT A\; \type$ and $\Gamma \vdashT B\; \type$ with $\eraseR{A} = \eraseR{B}$, then  $\Gamma \vdashT A = B$.

    \item If $\Gamma \vdashT t : \U_n$ and $\Gamma \vdashT A \; \type$ with $\eraseR{A} = \eraseR{t}$, then $\Gamma \vdashT A = T_n(t)$. 

    \item If $\Gamma \vdashT u_0 : A_0$ and $\Gamma \vdashT u_1 : A_1$ with $\eraseR{u_0} = \eraseR{u_1}$, then we have two cases:
    \begin{itemize}
        \item Either $\Gamma \vdashT A_0 = A_1$ and $\Gamma \vdashT u_0 = u_1 : A_0$.
        \item Or they are codes, such that $\Gamma \vdashT A_0 = \U_{n_0}$, $\Gamma \vdashT A_1 = \U_{n_1}$, and they agree
            at the join $k := \max(n_0, n_1)$ of their universes:
            \[
                \Gamma \vdashT t_{n_0}^{k}(u_0) = t_{n_1}^{k}(u_1) : \U_k.
            \]
    \end{itemize}
    \end{enumerate}
\end{lemma}

\begin{proof}
    We prove the property by mutual induction on $A$ and $u_0$, proving the three statements at the same time. For each of the cases of
    the induction, we need a "catch-up" lemma which inverts the erasure hypothesis. Concretely the issue is that we cannot 
    deduce from $\eraseR{u_1} = \Pi(A,B)$ that $u_1$ is a dependent product, as it could well be the decoding of a code $T_n(\Pi^n(a,b))$, or
    even a decoding of a lift $T_n(t_m^n(a))$. To solve this, we need to do another induction on $u_1$ for each case. We present key cases of the proof.

    \textit{Lambda abstraction.} If $u_0 = \lambda(C_0,B_0,b_0)$, then we necessarily have $\Gamma \vdashT A_0 = \Pi(C_0,B_0)$,
        and by inversion, the derivation $\Gamma \vdashT u_0 : A_0$ looks like:

    \begin{mathparsmall}
      \ebrule{
        \hypo{\Gamma.C_0 \vdashT b_0 : B_0}
        \infer1{\Gamma \vdashT \lambda(C_0,B_0,b_0) : \Pi(C_0,B_0)}
      }
    \end{mathparsmall}

    Then $\eraseR{u_0} = \lambda(\eraseR{C_0}, \eraseR{B_0}, \eraseR{b_0}) = \eraseR{u_1}$ and so $u_1$ is necessarily
    an abstraction or a lift. Now let's prove the property by induction on $u_1$.

    \begin{itemize}
        \item \underline{First case:} $u_1 = \lambda(C_1, B_1, b_1)$ with $\eraseR{C_1} = \eraseR{C_0}$, $\eraseR{B_1} = \eraseR{B_0}$, $\eraseR{b_1} = \eraseR{b_0}$.
    Also, by inversion $\Gamma \vdashT u_1 : A_1$ is necessarily of the same shape, with $\Gamma \vdashT A_1 = \Pi(C_1,B_1)$.
    Using the induction hypothesis on $C_0$ with $C_1$, we get $\Gamma \vdashT C_0 = C_1$.
    Then, by \zcref{Transform}, we get derivations $\Gamma.C_0 \vdashT B_1 \; \type$ and $\Gamma.C_0 \vdashT b_1 : B_1$.
    Applying the induction hypothesis on $B_0$ with $B_1$ yields $\Gamma.C_0 \vdashT B_0 = B_1$, and applying it on $b_0$ with $b_1$ gives two cases.

    In the first case, we directly get $\Gamma.C_0 \vdashT b_0 = b_1 : B_0$.

    In the second case, we get $\Gamma.C_0 \vdashT B_0 = \U_{n_0}$, $\Gamma.C_0 \vdashT B_1 = \U_{n_1}$, and
    $\Gamma.C_0 \vdashT t_{n_0}^{k}(b_0) = t_{n_1}^{k}(b_1) : \U_k$ where $k := \max(n_0, n_1)$.
    Since $\Gamma.C_0 \vdashT B_0 = B_1$, transitivity gives $\Gamma.C_0 \vdashT \U_{n_0} = \U_{n_1}$, hence $n_0 = n_1 = k$ by injectivity of universes,
    and therefore $\Gamma.C_0 \vdashT b_0 = b_1 : B_0$ using the identity lift equation $t_k^k(a) = a$.

    In both cases, using congruence equations for $\Pi$ and $\lambda$, we obtain the desired $\Gamma \vdashT A_0 = A_1$ and $\Gamma \vdashT u_0 = u_1 : A_0$.

    \item \underline{Second case:} $u_1 = t_n^m(a)$, and we will show that this case is not possible.
        By inversion, we have $\Gamma \vdashT A_1 = \U_m$ and $\Gamma \vdashT a : \U_n$.
        Since $\eraseR{u_1} = \eraseR{a} = \eraseR{u_0}$, we can apply the inner induction hypothesis on $a$,
        and in each case we obtain either $\Gamma \vdashT A_0 = \U_n$ or $\Gamma \vdashT A_0 = \U_{n_0}$,
        which contradict \zcref{no-confusion} since $\Gamma \vdashT A_0 = \Pi(C_0,B_0)$.
    \end{itemize}

    \textit{Code for a universe.} If $u_0$ is a code for a universe, then by inversion :
    \begin{mathparsmall}
      \ebrule{
        \hypo{}
        \infer1{\Gamma \vdashT \U_m^{n_0} : \U_{n_0}}
      }
    \end{mathparsmall}
    with $\Gamma \vdashT A_0 = \U_{n_0}$ and $m < n_0$. Then $\eraseR{u_0} = \U_m = \eraseR{u_1}$, so $u_1$ is either $\U_m^{n_1}$ or a lift $t_l^{n_1}(a)$. We proceed by induction on $u_1$.

    \begin{itemize}
        \item \underline{First case :} We have $u_1 = \U_m^{n_1}$, and by inversion $\Gamma \vdashT A_1 = \U_{n_1}$ with $m < n_1$.
    If we define $k := \max(n_0, n_1)$, then by the lifting conversion rule for universe codes we have
    $\Gamma \vdashT t_{n_0}^{k}(\U_m^{n_0}) = \U_m^k : \U_k$ and $\Gamma \vdashT t_{n_1}^{k}(\U_m^{n_1}) = \U_m^k : \U_k$,
    whence $\Gamma \vdashT t_{n_0}^{k}(u_0) = t_{n_1}^{k}(u_1) : \U_k$.

    \item \underline{Second case :} We have $u_1 = t_l^{n_1}(a)$ with $l \le n_1$, $\Gamma \vdashT a : \U_l$, and $\Gamma \vdashT A_1 = \U_{n_1}$.
    Since $\eraseR{a} = \eraseR{u_1} = \U_m$, the inner induction hypothesis on $a$ gives either $\Gamma \vdashT \U_{n_0} = \U_l$ and $\Gamma \vdashT u_0 = a : \U_{n_0}$,
    or $\Gamma \vdashT t_{n_0}^{k'}(u_0) = t_l^{k'}(a) : \U_{k'}$ for $k' := \max(n_0, l)$ (after identifying levels by injectivity of universes).
    In both cases, lifting to $k := \max(n_0, n_1)$ via congruence and functoriality of lifts yields
    $\Gamma \vdashT t_{n_0}^{k}(u_0) = t_{n_1}^{k}(u_1) : \U_k$.
    \end{itemize}
\end{proof}

\begin{lemma}[\formline{Equivalence.v}{11}{simplifyinduction}]\label{Utile}
     Assume that $\Gamma \vdashT u_0 : A_0$, $\Gamma \vdashT u_1 : A_1$, $\Gamma \vdashT A_0 = \U_{n_0}$, $\Gamma \vdashT A_1 = \U_{n_1}$, and
     $\Gamma \vdashT t_{n_0}^{k}(u_0) = t_{n_1}^{k}(u_1) : \U_k$ where $k := \max(n_0, n_1)$.

     Then if we additionally have $\Gamma \vdashT A_0 = A_1$, we get $\Gamma \vdashT u_0 = u_1: A_0$.
\end{lemma}
\begin{proof}
    By transitivity, we get $\Gamma \vdashT \U_{n_0} = \U_{n_1}$, from which we deduce $n_0 = n_1$ using
    injectivity of universes, and hence $k = n_0 = n_1$.
    Using the identity lift equation $t_{n_0}^{n_0}(a) = a$, we obtain $\Gamma \vdashT u_0 = u_1 : \U_{n_0}$ by transitivity,
    and thus $\Gamma \vdashT u_0 = u_1 : A_0$ by type conversion.
\end{proof}

Hence, we deduce the following corollary, which states that if we have two judgments
in \ThyTarski which are syntactically equal in \ThyRussell (when removing coercions, lifts...), then
they are provably convertible in \ThyTarski. This already means that the Tarski system is not much stronger than
the Russell one, in the sense that if two terms differ only by their explicit coercion, then they have the same
computational behaviour.

\begin{corollary}[\formline{Equivalence.v}{1128}{eraseinjtermplus} Judgemental injectivity of erasure]\label{uniqueness}
    If $\Gamma \vdashT u_0 : A_0$ and $\Gamma \vdashT u_1 : A_1$ in \ThyTarski and $\eraseR{u_0} = \eraseR{u_1}$,
    with the additional hypothesis $\eraseR{A_0} = \eraseR{A_1}$, then : $\Gamma \vdashT A_0 = A_1$ and 
    $\Gamma \vdashT u_0 = u_1 : A_0$.
\end{corollary}

\begin{proof}
    We apply \zcref{mainuniverse} on the equality $\eraseR{A_0} = \eraseR{A_1}$ to get $\Gamma \vdashT A_0 = A_1$, and on
    $\eraseR{u_0} = \eraseR{u_1}$ to get the two cases. In the first one, we directly have $\Gamma \vdashT u_0 = u_1 : A_0$,
    and in the second one, we use \zcref{Utile} to get the derivation.
\end{proof}

Now we prove the following theorem, stating the formal equivalence between the two systems:

\begin{theorem}[\agdalink{https://github.com/coquand/domain-semantics/blob/main/ERT/Equivalence.agda} \formline{Equivalence.v}{1379}{sectionctx} Section of erasure]\label{equT1T2}
    If $\Gamma \vdashR J$ in the system \ThyRussell, then there exists a unique, up to conversion judgment
    $\Gamma_1 \vdashT J_1$ in the system \ThyTarski such that $\eraseR{\Gamma_1} = \Gamma$ and $\eraseR{J_1} = J$.
\end{theorem}

\begin{proof}
    Uniqueness up to conversion follows directly from the previous lemma. We have to construct the derivation 
    by induction on the rules of \ThyRussell, making an extensive use of the previous lemma for rules with two premises. Indeed,
    when using the induction hypothesis on two premises, we need to make sure that the same types have 
    the same translations. To illustrate this, we detail the case of the product.

    \textit{Dependent Product.} If the last rule applied in \ThyRussell is :
    \begin{mathparsmall}
      \ebrule{
        \hypo{\Pi_1}
        \infer1{\Gamma \vdashR A \; \type}
        \hypo{\Pi_2}
        \infer1{\Gamma.A \vdashR B \; \type}
        \infer2{\Gamma \vdashR \Pi(A,B) \; \type}
      }
    \end{mathparsmall}

        Then, by the induction hypothesis we get derivations 
        $\Gamma' \vdashT A' \; \type$ and $\Gamma'',A'' \vdashT B' \; \type$ in \ThyTarski such that 
         $\eraseR{\Gamma'} = \eraseR{\Gamma''} = \Gamma$ and $\eraseR{A''} = \eraseR{A'} = A$. By \zcref{uniqueness}, we get $\Gamma' \vdashT A' = A''$, and  the contexts
        $\Gamma'$ and $\Gamma''$ are convertible. We can thus transform the second derivation to get $\Gamma',A' \vdashT B' \; \type$.
        Therefore, we can construct the derivation for $\Gamma' \vdashT \Pi(A',B') \; \type$ in \ThyTarski.
\end{proof}

This finally proves that any judgement in Russell-Style can be lifted uniquely to a judgement in Tarski-Style, proving
the equivalence between \ThyTarski and \ThyRussell by exhibiting a section of the forgetful map.

\begin{remark}
To prove this equivalence result, we could have relied on normalisation to simplify the proofs, at
the cost of relying on a stronger metatheory and stronger properties of the theory. In fact, if we proved normalisation for
\ThyRussell as well, we could construct the derivation of \zcref{equT1T2} by an induction on normal forms.
Then, by showing that the erasure map $\eraseR{\cdot}$ commutes with $\nf(.)$, we would obtain the equivalence result.
\end{remark}

\subsection{Unannotated syntax (\ThyRussell and \ThyPTS)}\label{sectionT2T2'}

Finally, we prove our last equivalence result between \ThyRussell and \ThyPTS, relating
annotated and unannotated syntax. Concretely, the target theory \ThyPTS is the closest we can
get to a presentation as a Pure Type System of Martin-Löf's type theory with a cumulative hierarchy of universes.
This section is a direct adaptation of the arguments of \citet{Streicher} to our setting. Notably, we need
to adapt the arguments because the theory \ThyRussell does not enjoy uniqueness of types, precisely because
of cumulativity.

Just as we did for the previous equivalence result, we prove that the forgetful map $\eraseP{\cdot}$ preserves
derivations.

\begin{theorem}[\agdalink{https://github.com/coquand/domain-semantics/blob/main/ERT/StripDeriv.agda} Correctness]
    If $\Gamma \vdashR J$ in \ThyRussell, then we have a derivation $\eraseP{\Gamma} \vdashP \eraseP{J}$.
\end{theorem}
\begin{proof}
    We just need to prove that the stripping of rules in \ThyRussell are admissible in \ThyPTS, which
    is quite easy as each rule in \ThyRussell corresponds directly to a rule of \ThyPTS. Then, the theorem
    is proved by an induction on the derivation which replaces rules in \ThyRussell by the admissible ones in \ThyPTS.
\end{proof}

Now, following the ideas of \citet{Streicher}, we want to prove that two terms or types which
become syntactically equal after erasing the annotations with $\eraseP{\cdot}$ are convertible in \ThyRussell. Here, 
types are not unique because of cumulativity, so we could well have $\eraseP{u_0} = \eraseP{u_1}$ without
necessarily having $u_0$ and $u_1$ convertible. This is due to the codomain annotations of $\lambda$-abstractions, which are erased by $\eraseP{\cdot}$ and are subject to cumulativity, as shown by the following counter-example.

\begin{example}\label{ex:cumul-lam}
    For $n < p$, the terms $f_0 = \lambda(\U_n,\U_n,v_0) : \Pi(\U_n,\U_n)$ and $f_1 = \lambda(\U_n,\U_p,v_0) : \Pi(\U_n,\U_p)$ have the same stripping $\lambda(\U_n,v_0)$,
    but neither their types $\Pi(\U_n,\U_n)$ and $\Pi(\U_n,\U_p)$ nor $f_0$ and $f_1$ are convertible.
    Yet, when applied to $\U_0$ (for $0 < n < p$), the applications $u_0 = \app(\U_n,\U_n,f_0,\U_0)$ and $u_1 = \app(\U_n,\U_p,f_1,\U_0)$ both have type $\U_p$ by cumulativity, have the same stripping $(\lambda(\U_n,v_0))\;\U_0$, and are convertible: $\Gamma \vdashR u_0 = u_1 : \U_p$.
\end{example}

As \zcref{ex:cumul-lam} illustrates, even if two applications $u_0$ and $u_1$ are compared at a common type, the functions $f_0$ and $f_1$ being applied may have types $\Pi(\U_n,\U_n)$ and $\Pi(\U_n,\U_p)$ that differ by cumulativity \emph{under $\Pi$}.
Although \ThyRussell has no primitive sub-typing judgement for $\Pi$-types, we can relate such terms by defining explicit coercions into a common upper type (a \emph{join}) using $\eta$-expansion, with cumulativity acting at the leaves.

\begin{definition}[Joins and coercions]\label{def:join}
    We define $\Gamma \vdashR \JoinTy(T_0, T_1, L) \leadsto (\coe_0, \coe_1)$, relating types $T_0, T_1$ to a common join $L$ via coercion operations $\coe_0, \coe_1$, by the following  rules:
    \begin{mathparsmall}
      \ebrule{
        \hypo{\Gamma \vdashR T_0 = L}
        \hypo{\Gamma \vdashR T_1 = L}
        \infer2{\Gamma \vdashR \JoinTy(T_0, T_1, L) \leadsto (\Id, \Id)}
      }
      \and
      \ebrule{
        \hypo{\Gamma \vdashR T_0 = \U_k}
        \hypo{\Gamma \vdashR T_1 = \U_m}
        \infer2{\Gamma \vdashR \JoinTy(T_0, T_1, \U_{\max(k,m)}) \leadsto (\Id, \Id)}
      }
      \and
      \ebrule{
        \hypo{\Gamma \vdashR T_0 = \Pi(A, B_0)}
        \hypo{\Gamma \vdashR T_1 = \Pi(A, B_1)}
        \hypo{\Gamma.A \vdashR \JoinTy(B_0, B_1, L') \leadsto (\coe_0, \coe_1)}
        \infer3{\Gamma \vdashR \JoinTy(T_0, T_1, \Pi(A, L')) \leadsto (\pi_{A,B_0,L'}(\coe_0), \pi_{A,B_1,L'}(\coe_1))}
      }
    \end{mathparsmall}
    where the action of a coercion on a term $t$ is defined inductively by:
    \begin{align*}
        \Id(t) &:= t,\\
        \pi_{A,B,L'}(\coe)(t) &:= \lambda(A, L', \coe(\app(A\p, B\p^+, t\p, v_0))).
    \end{align*}
\end{definition}

We can prove that joins are symmetric, stable under type conversion and substitution. Coercions are well-typed, congruent, and compute under application by the $\beta$-rule:
\[\Gamma \vdashR \app(A, L', \pi_{A,B_i,L'}(\coe_i)(t), a) = \coe_{i[a]}(\app(A, B_i, t, a)) : L'[a].\]

When $T_0$ and $T_1$ are convertible, coercions into their join are injective up to conversion (analogously to \zcref{Utile}).

\begin{lemma}[Diagonal lemma]\label{lem:join-diag}
    If $\Gamma \vdashR \JoinTy(T_0, T_1, L) \leadsto (\coe_0, \coe_1)$, $\Gamma \vdashR t_0 : T_0$, $\Gamma \vdashR t_1 : T_1$, and $\Gamma \vdashR \coe_0(t_0) = \coe_1(t_1) : L$, then $\Gamma \vdashR T_0 = T_1$ implies $\Gamma \vdashR t_0 = t_1 : T_0$.
\end{lemma}
\begin{proof}
    By induction on $\Gamma \vdashR \JoinTy(T_0, T_1, L) \leadsto (\coe_0, \coe_1)$. The conversion case is immediate.

    \textit{Universe case.} $\Gamma \vdashR \U_k = \U_m$ gives $k = m$ by universe injectivity, so $L = \U_{\max(k,k)} = \U_k$.

    \textit{Product case.} $\Gamma \vdashR \Pi(A, B_0) = \Pi(A, B_1)$ gives $\Gamma.A \vdashR B_0 = B_1$ by $\Pi$-injectivity. Weakening $\Gamma \vdashR \pi_{A,B_0,L'}(\coe_0)(t_0) = \pi_{A,B_1,L'}(\coe_1)(t_1) : \Pi(A, L')$ to $\Gamma.A$, applying both sides to $v_0$, and using the $\beta$-rule gives $\Gamma.A \vdashR \coe_0(h_0) = \coe_1(h_1) : L'$ where $h_i = \app(A\p, B_i\p^+, t_i\p, v_0)$. The induction hypothesis gives $\Gamma.A \vdashR h_0 = h_1 : B_0$, and congruence of $\lambda$ together with the $\eta$-rule concludes $\Gamma \vdashR t_0 = t_1 : T_0$.
\end{proof}

We now state and prove the main lemma of this section by structural induction on the syntax, comparing terms at the join of their types.

\begin{lemma}\label{streichT2T2'}
    In \ThyRussell, we have:
    \begin{enumerate}
        \item If $\Gamma \vdashR u_0 : T_0$ and $\Gamma \vdashR u_1 : T_1$ with $\eraseP{u_0} = \eraseP{u_1}$, then there exists a join $\Gamma \vdashR \JoinTy(T_0, T_1, L) \leadsto (\coe_0, \coe_1)$ such that $\Gamma \vdashR \coe_0(u_0) = \coe_1(u_1) : L$.
        \item If $\Gamma \vdashR A_0 \; \type$ and $\Gamma \vdashR A_1 \; \type$ with $\eraseP{A_0} = \eraseP{A_1}$, then $\Gamma \vdashR A_0 = A_1$.
    \end{enumerate}
\end{lemma}

\begin{proof}
    By mutual induction on $u_0$ and $A_0$. Statement (2) follows directly from statement (1), since for $\Gamma \vdashR A_i : \U_{l_i}$, any join of $\U_{l_0}$ and $\U_{l_1}$ gives $\Gamma \vdashR A_0 = A_1 : \U_{\max(l_0, l_1)}$ by no-confusion and universe injectivity.
    For statement (1), inversion on $\Gamma \vdashR u_i : T_i$ gives the principal type $P_i$ of $u_i$, which is related to $T_i$ by conversion or cumulativity. It therefore suffices to construct a join for the principal types $P_0, P_1$.

    Since $\eraseP{u_0} = \eraseP{u_1}$, $u_0$ and $u_1$ have the same head constructor. Variables, universes, and products are straightforward. For $\lambda$-abstractions $u_i = \lambda(A_i, B_i, b_i)$, statement (2) gives $\Gamma \vdashR A_0 = A_1$. After transporting $b_1 : B_1$ to context $\Gamma.A_0$, the induction hypothesis on $b_0, b_1$ gives a join $\Gamma.A_0 \vdashR \JoinTy(B_0, B_1, L') \leadsto (\coe_0, \coe_1)$ with $\Gamma.A_0 \vdashR \coe_0(b_0) = \coe_1(b_1) : L'$, which lifts to a product join on $\Pi(A_0, L')$ via the $\beta$-rule under $\lambda$. We detail the remaining case of function application.

    \textit{Function application.} If $u_0 = \app(C_0, D_0, c_0, a_0)$, then $u_1 = \app(C_1, D_1, c_1, a_1)$ with $\eraseP{c_0} = \eraseP{c_1}$ and $\eraseP{a_0} = \eraseP{a_1}$. By inversion, $\Gamma \vdashR c_i : \Pi(C_i, D_i)$ and $\Gamma \vdashR a_i : C_i$, with principal type $P_i = D_i[a_i]$.
    By the induction hypothesis, we have a join $\Gamma \vdashR \JoinTy(\Pi(C_0, D_0), \Pi(C_1, D_1), L) \leadsto (\coe_0, \coe_1)$ with $\Gamma \vdashR \coe_0(c_0) = \coe_1(c_1) : L$, and a join for $a_0, a_1$ on $C_0, C_1$. By no-confusion, the join for $c_0, c_1$ is either a conversion join or a product join:
    \begin{itemize}
        \item \underline{Conversion join:} We have $\Gamma \vdashR \Pi(C_0, D_0) = \Pi(C_1, D_1)$ and $\Gamma \vdashR c_0 = c_1 : \Pi(C_0, D_0)$. By $\Pi$-injectivity, $\Gamma \vdashR C_0 = C_1$ and $\Gamma.C_0 \vdashR D_0 = D_1$. \zcref{lem:join-diag} on the join for $a_0, a_1$ gives $\Gamma \vdashR a_0 = a_1 : C_0$, and congruence of $\app$ concludes.
        \item \underline{Product join:} We have $L = \Pi(A, L')$ with $\Gamma \vdashR \Pi(C_i, D_i) = \Pi(A, B_i)$ and $\Gamma.A \vdashR \JoinTy(B_0, B_1, L') \leadsto (\coe_{B_0}, \coe_{B_1})$. By $\Pi$-injectivity, $\Gamma \vdashR C_i = A$ and $\Gamma.C_i \vdashR D_i = B_i$. In particular, $\Gamma \vdashR C_0 = C_1$, so \zcref{lem:join-diag} again gives $\Gamma \vdashR a_0 = a_1 : A$. Applying $\Gamma \vdashR \coe_0(c_0) = \coe_1(c_1) : \Pi(A, L')$ to $a_0$, using the $\beta$-computation rule for coercions on both sides, and congruence of $\app$ gives: 
        $\Gamma \vdashR \coe_{B_0[a_0]}(u_0) = \coe_{B_1[a_0]}(u_1) : L'[a_0]$
        which is a join for $B_0[a_0]$ and $B_1[a_0]$, and hence for $D_0[a_0]$ and $D_1[a_1]$.
    \end{itemize}
\end{proof}

Combining \zcref{streichT2T2'} with the diagonal lemma (\zcref{lem:join-diag}), we deduce the following direct corollary, analogous to \zcref{uniqueness} of the previous section.

\begin{corollary}[\agdalink{https://github.com/coquand/domain-semantics/blob/main/ERT/StripUniqNF.agda}]\label{cor:strip-uniq}
    If $\Gamma \vdashR u_0 : A_0$ and $\Gamma \vdashR u_1 : A_1$ in \ThyRussell with $\eraseP{u_0} = \eraseP{u_1}$ and $\eraseP{A_0} = \eraseP{A_1}$, then $\Gamma \vdashR A_0 = A_1$ and $\Gamma \vdashR u_0 = u_1 : A_0$.
\end{corollary}
\begin{proof}
    If $\eraseP{A_0} = \eraseP{A_1}$, point (2) of \zcref{streichT2T2'} gives $\Gamma \vdashR A_0 = A_1$. Then point (1) and \zcref{lem:join-diag} give $\Gamma \vdashR u_0 = u_1 : A_0$.
\end{proof}

Finally, we can show the following theorem, which exhibits the section of the forgetful map $\eraseP{\cdot}$, showing the equivalence
of \ThyRussell and \ThyPTS.

\begin{theorem}[\agdalink{https://github.com/coquand/domain-semantics/blob/main/ERT/SectionR.agda}]\label{equT2T2'}
    If $\Gamma \vdashP J$ in \ThyPTS, then we have a unique up to conversion judgment 
    $\Gamma' \vdashR J'$ in \ThyRussell such that $\eraseP{\Gamma'} = \Gamma$ and $\eraseP{J'} = J$.
\end{theorem}
\begin{proof}
    By induction on the derivation $\Gamma \vdashP J$, using \zcref{cor:strip-uniq}, we construct a derivation in \ThyRussell.
    The main point is that the typing annotations can be inferred from the derivation, and that \zcref{cor:strip-uniq} gives
    us uniqueness up to conversion for rules with shared terms or types in their premises. 
\end{proof}

In \ThyPTS, we could also remove the judgement $\Gamma \vdashP A\; \type$ to manipulate only universes,
in a completely Russellian style. In such a system, there is no top-level of types and a type is simply a term
of a certain universe.
We can also consider named variables instead of de Bruijn indices, to get a theory presented exactly like a pure type
system, with the additional rule for cumulativity.

If we wanted to go even further, we could also remove the annotation of the variable from the lambda abstraction, again
following \citet{Streicher}.
However to do so, we crucially need to rely on the full normalisation result, as removing this annotation breaks the structural induction of \zcref{streichT2T2'}: the same judgement can be derived using subterms with non-convertible types, as shown by the following example.
    
\begin{example}
        We have two derivations: $t :\U_n \vdashP \lambda \,x.x : \Pi(\U_n,\U_n)$ and $t :\U_n \vdashP \lambda \,x.x : \Pi(\U_n,\U_p)$, whose types are not convertible when $n<p$.
        Then we can derive the same judgement from those two premises: $t :\U_n \vdashP (\lambda \,y.t) \; (\lambda \,x.x) : \U_n$.
        While the two resulting derivations in \ThyRussell are still convertible (both $\beta$-reduce to $t$), the two lifts of $\lambda \,y.t$ have non-convertible domain annotations $\Pi(\U_n,\U_n)$ and $\Pi(\U_n,\U_p)$ and therefore admit no join, preventing a direct proof by structural induction without normalisation.
\end{example}

\section{Semantics and Initiality for implicit syntax}\label{sec:semantics}

We now show how the syntactic equivalences of Sections~\ref{sectionT0T1}--\ref{sectionT2T2'} allow us to interpret the implicit presentation \ThyPTS{} in semantic models of type theory, and to extend initiality.

While we used the usual presentation with pairing $(\sigma, a)$ for \ThyGAT, giving interpretation in models is simpler using substitutions à la Ehrhard \cite{Ehrhard1988}. We define a variant \ThyE{} of \ThyGAT{} with operations $\cdot^+$ and $[\cdot]$:

\begin{definition}[Ehrhard's substitutions]
    The theory \ThyE{} is obtained from \ThyGAT{} by replacing pairing $(\cdot,\cdot)$ with operations $\cdot^+$ and $[\cdot]$:
    \begin{align*}
        \Gamma, \Delta : \ctx, A : \Type(\Gamma), \sigma : \Delta \rightarrow \Gamma &\vdashE \sigma^{+(A)} : \Delta.A\sigma \rightarrow \Gamma.A \\
        \Gamma : \ctx, A : \Type(\Gamma), a : \Elem(\Gamma, A) &\vdashE [a]_A : \Gamma \rightarrow \Gamma.A
    \end{align*}
    \begin{gather*}
        \Id^{+(A)} = \Id, \qquad
        (\sigma\delta)^{+(A)} = \sigma^{+(A)}\,\delta^{+(A\sigma)}, \qquad
        \p\,\sigma^{+(A)} = \sigma\,\p, \qquad
        \q\,\sigma^{+(A)} = \q, \\
        \p\,[a]_A = \Id, \qquad
        \q\,[a]_A = a, \qquad
        [a]_A\,\sigma = \sigma^{+(A)}\,[a\sigma]_{A\sigma}, \qquad
        \p^{+(A)}\,[\q]_{A\p} = \Id.
    \end{gather*}
\end{definition}

This presentation makes formalisation of models easier in an intensional metatheory, as the laws of usual type formers such as $\Pi$, $\Sigma$ use those primitives, whereas with pairing we would need to transport equality between types in the model.

\begin{lemma}[Equivalence of \ThyGAT{} and \ThyE{}]\label{lem:gat-ehrhard}
    The theories \ThyGAT{} and \ThyE{} are equivalent: their substitution operations are interdefinable and mutually inverse up to conversion.
\end{lemma}
\begin{proof}
    We translate \ThyE{} into \ThyGAT{} by defining $\sigma^{+(A)} := (\sigma\p, \q)$ and $[a]_A := (\Id, a)$. All equations follow from the pairing laws of \ThyGAT{}. Conversely, we define $(\sigma, a) := \sigma^{+(A)}[a]_{A\sigma}$ and again all equations are directly derivable, for instance:
    $\p(\sigma, a) = \p\,\sigma^{+(A)}\,[a]_{A\sigma} = \sigma\,\p\,[a]_{A\sigma} = \sigma$.
    Both round-trips reduce to the identity up to conversion.
\end{proof}

Any derivation is finite and only refers to a finite number of universe levels. Hence, we will interpret them in models with a bounded universe hierarchy.

\begin{definition}[Model]
    A \emph{model of level $N$} is a category with families in Ehrhard's presentation, equipped with dependent products and a cumulative Tarski hierarchy $\U_0, \dots, \U_N$ with decoding $T_n$, codes $\Pi^n, \U_m^n$, and lifts $t_m^n$.
\end{definition}

We can define an interpretation for expressions of \ThyE{} in a given model $\mathcal{M}$ \cite{Hofmann1997, CWF-CCD, noteGAT}. We can prove soundness for this interpretation by induction, which composes with our equivalences ($\ThyPTS \simeq \ThyRussell \simeq \ThyTarski \simeq \ThyGAT \simeq \ThyE$) to give the following interpretation result.

\begin{theorem}[Semantics of \ThyPTS{}]\label{thm:russell-semantics}
    For every model $\mathcal{M}$ of level $N \in \bN$, every derivable judgment of \ThyPTS{} bounded by level $N$ has a unique denotation in $\mathcal{M}$ via \ThyE{}, independently of the choice of lift, and such that convertible contexts, types, and terms have equal denotations.
\end{theorem}
\begin{proof}
    Existence follows by lifting a judgment along our equivalences and soundness of the semantics. For uniqueness, two lifts of the same judgment have convertible contexts, lifted types and terms in \ThyE{}, so by soundness they have the same denotation. Finally, lifting a conversion judgment in \ThyPTS{} gives an equality in \ThyE{}, with a common denotation.
\end{proof}

A morphism of models consists of maps on contexts, substitutions, types, and terms which preserve every operation. The initial model for this notion of morphism is the term model $\mathcal{I}_N$ of \ThyE{} (at a level $N \in \bN$), whose types and terms correspond to the syntactic types and terms of \ThyE{}, paired with their contexts and quotiented by the equations \cite{Cartmell, CWF-CCD, Brunerie}.

\begin{theorem}[Initiality]\label{thm:initiality}
    For every $N \in \bN$, $\mathcal{I}_N$ is initial i.e. for every model $\mathcal{M}$ of level $N$, there exists a unique morphism $\mathcal{I}_N \to \mathcal{M}$, given by the interpretation map.
\end{theorem}
\begin{proof}
    Existence is obtained by interpreting representatives, and soundness guarantees that it does not depend on this choice. For uniqueness, an induction shows that two morphisms $\mathcal{I}_N \to \mathcal{M}$ agree on the class of every derivable context, type, term, and substitution.
\end{proof}

This means that through our equivalences, the elements of $\mathcal{I}_N$ correspond bijectively to derivable judgments of \ThyPTS{} modulo conversion. In the formalisation \cite{unit-formalisation}, we instantiated \zcref{thm:russell-semantics} in three concrete models.
\begin{itemize}
    \item \emph{Iterative sets.} Following \cite{Gylterud2018, aczel}, we interpret types as families of iterative sets and $\U_n$ is the set $V_n$ of iterative sets at meta-level $n$. Here the universe lift $V_m \to V_n$ is not the identity. This illustrates that it is crucial to interpret from a theory with Tarski universes, as a direct interpretation of \ThyPTS{} would otherwise have to choose lifts and prove coherence.
    \item \emph{Univalent posets.} For $N=0$, we give an interpretation in the category with families of fibrant types on finite posets \cite{Coquand2026Unit} (with fibrations given by relative contractibility), equipped with one universe $\U_0$ closed under $\Pi$ with strict decoding $T(\ulcorner A \urcorner) = A$, natural in substitutions.
    \item \emph{Simplicial sets.} On simplicial sets, the classifier consists of horn-filling operations, and the fibration structure of $\mathsf{El}$ is obtained by turning horn fillers into box fillers and back. We take two universes $\U_0 \subset \U_1$ of horn-filling operations, with inclusion $\iota : \U_0 \rightarrow \U_1$, and the code-map category with families: a type over $\Gamma$ is a map $A : \Gamma \rightarrow \U_1$, its terms are the sections of $\mathsf{El}_1$ pulled back along $A$. The decoding is defined as $T(a) := \iota \circ a$ and its strictness follows from horn-to-box filling commuting along monomorphisms.
\end{itemize}

\section{Extension to Explicit Universe Polymorphism}\label{extension}

In this section, we illustrate the modularity of our proof method by showing that it extends to the theory of \citet{polymorphism, bcde2026}, with explicit universe polymorphism and cumulative universes indexed by internal levels.

For the theory without constraints, the proof of the previous sections can be adapted directly by simply making the reasoning on levels internal to the theory. We already know that this theory enjoys definitional inversion as it is normalising \cite{fuss-free}. However, for a theory with constraints, things get more complicated. Here, we only sketch the proof idea, but the proof for this extended system has been auto-formalised in Agda \cite{auto-formalisation}.

In this system, levels are now an actual sort with variables, suprema, and successors: 
$l ::= \alpha_i \mid l \vee m \mid l^+$
Additionally, we have constraints of the form $\psi ::= l = m$, which can represent $l \leq m$ as $l \vee m = m$ and $l < m$ as $l^+ \leq m$.
Contexts can be extended with level variables $\Gamma, \alpha$ and constraints $\Gamma, \psi$. We add level and constraints products $[\alpha]A$, $[\psi]A$, together with an empty type $\emptyset : \U_l$.
Those new constructs obey to the following rules. 

\begin{mathparsmall}
  \ebrule{
    \hypo{\Gamma, \alpha \vdash A \; \type}
    \infer1{\Gamma \vdash [\alpha]A \; \type}
  }
  \and
  \ebrule{
    \hypo{\Gamma, \alpha \vdash u : A}
    \infer1{\Gamma \vdash \langle \alpha \rangle_A u : [\alpha]A}
  }
  \and
  \ebrule{
    \hypo{\Gamma \vdash t : [\alpha]A}
    \infer1{\Gamma \vdash t_A l : A[l/\alpha]}
  }
  \and
  \ebrule{
    \hypo{\Gamma, \alpha \vdash u : A}
    \infer1{\Gamma \vdash (\langle \alpha \rangle_A u)_A l = u[l/\alpha] : A[l/\alpha]}
  }
  \and
  \ebrule{
    \hypo{\Gamma \vdash t : [\alpha]A}
    \infer1{\Gamma \vdash t = \langle \alpha \rangle_A (t^\uparrow_{A^\uparrow}\,\alpha) : [\alpha]A}
  }
  \and
  \ebrule{
    \hypo{\Gamma, \psi \vdash A \; \type}
    \infer1{\Gamma \vdash [\psi]A \; \type}
  }
  \and
  \ebrule{
    \hypo{\Gamma, \psi \vdash t : A}
    \infer1{\Gamma \vdash \langle \psi \rangle_A t : [\psi]A}
  }
  \and
  \ebrule{
    \hypo{\psi \text{ valid in } \Gamma}
    \hypo{\Gamma \vdash A \; \type}
    \infer2{\Gamma \vdash [\psi]A = A}
  }
  \and
  \ebrule{
    \hypo{\psi \text{ valid in } \Gamma}
    \hypo{\Gamma \vdash t : A}
    \infer2{\Gamma \vdash \langle \psi \rangle_A t = t : A}
  }
\end{mathparsmall}

\noindent In addition, constraint products respect constraint equivalence $\psi \Leftrightarrow \psi'$, and whenever the level context has a \emph{loop} (i.e., $l < l$ is valid), collapse rules equate every type $A$ to $\emptyset$ and every term $t : A$ to $\emptyset : A$.

First, we relate the Tarski and Russell presentations (\ThyTarski and \ThyRussell). Because \zcref{mainuniverse} compares codes by lifting them to the join of their universes, the proof of \zcref{sectionT1T2} adapts directly by replacing $\max(n_0, n_1)$ with the internal supremum $n_0 \vee n_1$.
To handle inconsistent level contexts, we use the decidability of loop-checking \cite{loopchecking}: at each step of the proof, we first decide whether $\Gamma$ has a loop. If it doesn't, universe injectivity holds and the proof of \zcref{sectionT1T2} goes through, and otherwise the statement holds vacuously.

Second, we relate the annotated and unannotated presentations (\ThyRussell and \ThyPTS), where $\eraseP{\cdot}$ also erases the type annotations on $\langle \alpha \rangle_A u$, $\langle \psi \rangle_A t$, and $t_A l$. We extend the join and coercion method of \zcref{sectionT2T2'}. Because cumulativity does not act structurally on $[\alpha]A$ or $[\psi]A$, we extend \zcref{def:join} with join rules for $[\alpha]B_i$ and $[\psi]B_i$, where universe joins use the internal supremum $\U_{k \vee m}$ and the corresponding coercions are defined by $\eta$-expansion:
\[
    \coe_{[\alpha]B \leq [\alpha]L'}(t) := \langle \alpha \rangle_{L'}\,\coe_{B \leq L'}(t^\uparrow_{B^\uparrow}\,\alpha), \qquad
    \coe_{[\psi]B \leq [\psi]L'}(t) := \langle \psi \rangle_{L'}\,\coe_{B \leq L'}(t).
\]
For the diagonal lemma (\zcref{lem:join-diag}) to go through for constraint products, we add the natural $\eta$-rule for constraint products:
\[
    \Gamma \vdashR \langle \psi \rangle_A t = t : [\psi]A \quad \text{when} \quad \Gamma \vdashR t : [\psi]A.
\]

Moreover, since $[\psi]A = A$ holds whenever $\psi$ is valid in $\Gamma$, a constraint product is not necessarily distinct from $\U_l$, $\Pi$, or $[\alpha]A$.
Thus, we need a \emph{guard-aware} no-confusion property:

\begin{theorem}[Guard-aware no-confusion]
In a loop-free context $\Gamma$, if $\Gamma \vdashR X = [\psi]A$ where $X$ is $\U_l$, $\Pi(C,B)$, or $[\alpha]B$, then $\psi$ must be valid in $\Gamma$.
\end{theorem}
\begin{proof}
This follows directly from soundness of domain model. In this model, whenever $\psi$ is not valid in $\Gamma$, the type $[\psi]A$ denotes the trivial singleton $\{\bot\}$, whereas $\U_l$, $\Pi(C,B)$, and $[\alpha]B$ all contain non trivial elements. Hence, any conversion $\Gamma \vdashR X = [\psi]A$ forces $\psi$ to be valid in $\Gamma$.
\end{proof}

In the join proof, whenever a join is inverted against $\U_l$, $\Pi$, or $[\alpha]$, this property guarantees that any outer guard joins have valid constraints and can be stripped using $[\psi]A = A$.

\section{Conclusion and Future Work}

We proved that the Generalised Algebraic Theory and Pure Type System
presentations of Martin Löf's Type Theory are equivalent. In particular, we related 
\textit{Tarski} and \textit{Russell} presentations of cumulative universe hierarchies, and extended our equivalence result to universe polymorphism.
Those equivalence results are fundamental to justify the common, but informal, practice of using implicit syntax and Russell universes. Notably, our result rigorously legitimates the interpretation of implicit presentations in the usual models of type theory, which usually apply to fully annotated syntax.

Our proof relies solely on definitional inversion properties, which we prove using the domain-theoretic approach of \citet{definitional-inv}.  Crucially, we do not rely on normalisation and our proof is completely elementary. We believe that our proof technique can be adapted to a variety of extensions and more sophisticated type theories.

In this paper, we did not consider cumulativity with a structural sub-typing relation for the sake of simplicity, as it could lead to issues with general inductive types as pointed out by \citet{Luo}. Yet, structural sub-typing would provide a more convenient manipulation of universes in a practical proof assistant. The recent work of \citet{AdapTT} presents a general framework which provides all the necessary equations, and we hope that the results of this paper could be extended to this setting.

\bibliography{main}

\appendix

\section{Detailed rules}\label{appendix:rules}

All systems share the same rules for context well-formation and context conversion:

\begin{mathparsmall}
  \ebrule{
    \hypo{}
    \infer1{1 \vdash}
  }
  \and
  \ebrule{
    \hypo{\Gamma \vdash}
    \hypo{\Gamma \vdash A \;\type}
    \infer2{\Gamma.A \vdash}
  }
  \and
  \ebrule{
    \hypo{}
    \infer1{\vdash 1 = 1}
  }
  \and
  \ebrule{
    \hypo{\vdash \Gamma = \Delta}
    \hypo{\Gamma \vdash A = B}
    \infer2{\vdash \Gamma.A = \Delta.B}
  }
\end{mathparsmall}

\subsection{The system \ThyTarski : Tarski universes}

\begin{mathparsmall}
  \ebrule{
    \hypo{\Gamma \vdashT A \;\type}
    \infer1{\Gamma.A \vdashT v_0 : A\p}
  }
  \and
  \ebrule{
    \hypo{\Gamma \vdashT v_n : A}
    \hypo{\Gamma \vdashT B \;\type}
    \infer2{\Gamma.B \vdashT v_{n+1} : A\p}
  }
  \and
  \ebrule{
    \hypo{\Gamma \vdashT A \;\type}
    \hypo{\Gamma.A \vdashT B \;\type}
    \infer2{\Gamma \vdashT \Pi(A, B) \;\type}
  }
  \and
  \ebrule{
    \hypo{\Gamma.A \vdashT b : B}
    \hypo{\Gamma.A \vdashT B \;\type}
    \hypo{\Gamma \vdashT A \;\type}
    \infer3{\Gamma \vdashT \lambda(A,B,b) : \Pi(A, B)}
  }
  \and
  \ebrule{
    \hypo{\Gamma \vdashT c : \Pi(A, B)}
    \hypo{\Gamma \vdashT a : A}
    \hypo{\Gamma.A \vdashT B \;\type}
    \hypo{\Gamma \vdashT A \;\type}
    \infer4{\Gamma \vdashT \app(A,B,c,a) : B[a]}
  }
  \and
  \ebrule{
    \hypo{\Gamma \vdashT}
    \infer1{\Gamma \vdashT \U_n \;\type}
  }
  \and
  \ebrule{
    \hypo{\Gamma \vdashT a : \U_n}
    \infer1{\Gamma \vdashT T_n(a) \;\type}
  }
  \and
  \ebrule{
    \hypo{\Gamma \vdashT a : \U_n}
    \hypo{\Gamma.T_n(a) \vdashT b : \U_n}
    \hypo{\Gamma \vdashT T_n(a) \;\type}
    \infer3{\Gamma \vdashT \Pi^n(a, b) : \U_n}
  }
  \and
  \ebrule{
    \hypo{\Gamma \vdashT}
    \infer1[($m<n$)]{\Gamma \vdashT \U_m^n : \U_n}
  }
  \and
  \ebrule{
    \hypo{\Gamma \vdashT a : \U_m}
    \infer1[($m\leq n$)]{\Gamma \vdashT t_m^n(a) : \U_n}
  }
  \and
  \ebrule{
    \hypo{\Gamma \vdashT a : A}
    \hypo{\Gamma \vdashT A = B}
    \infer2{\Gamma \vdashT a : B}
  }
  \and
  \ebrule{
    \hypo{\Gamma \vdashT a = a' : A}
    \hypo{\Gamma \vdashT A = B}
    \infer2{\Gamma \vdashT a = a' : B}
  }
  \and
  \ebrule{
    \hypo{\Gamma \vdashT A \;\type}
    \infer1{\Gamma \vdashT A = A}
  }
  \and
  \ebrule{
    \hypo{\Gamma \vdashT A = B}
    \hypo{\Gamma \vdashT A = C}
    \infer2{\Gamma \vdashT B = C}
  }
  \and
  \ebrule{
    \hypo{\Gamma \vdashT a : A}
    \infer1{\Gamma \vdashT a = a : A}
  }
  \and
  \ebrule{
    \hypo{\Gamma \vdashT a = b : A}
    \hypo{\Gamma \vdashT a = c : A}
    \infer2{\Gamma \vdashT b = c : A}
  }
  \and
  \ebrule{
    \hypo{\Gamma \vdashT A = A'}
    \hypo{\Gamma.A \vdashT B = B'}
    \infer2{\Gamma \vdashT \Pi(A, B) = \Pi(A', B')}
  }
  \and
  \ebrule{
    \hypo{\Gamma \vdashT A = A'}
    \hypo{\Gamma.A \vdashT B = B'}
    \hypo{\Gamma.A \vdashT b = b' : B}
    \infer3{\Gamma \vdashT \lambda(A,B,b) = \lambda(A',B',b') : \Pi(A, B)}
  }
  \and
  \ebrule{
    \hypo{\Gamma \vdashT A = A'}
    \hypo{\Gamma.A \vdashT B = B'}
    \hypo{\Gamma \vdashT c = c' : \Pi(A, B)}
    \hypo{\Gamma \vdashT a = a' : A}
    \infer4{\Gamma \vdashT \app(A,B,c,a) = \app(A',B',c',a') : B[a]}
  }
  \and
  \ebrule{
    \hypo{\Gamma \vdashT a = a' : \U_n}
    \infer1{\Gamma \vdashT T_n(a) = T_n(a')}
  }
  \and
  \ebrule{
    \hypo{\Gamma \vdashT a = a' : \U_n}
    \infer1[($n\leq p$)]{\Gamma \vdashT t_n^p(a) = t_n^p(a') : \U_p}
  }
  \and
  \ebrule{
    \hypo{\Gamma \vdashT a = a' : \U_n}
    \hypo{\Gamma.T_n(a) \vdashT b = b' : \U_n}
    \infer2{\Gamma \vdashT \Pi^n(a, b) = \Pi^n(a', b') : \U_n}
  }
  \and
  \ebrule{
    \hypo{\Gamma.A \vdashT b : B}
    \hypo{\Gamma \vdashT a : A}
    \infer2[($\beta$-rule)]{\Gamma \vdashT \app(A,B,\lambda(A,B,b),a) = b[a] : B[a]}
  }
  \and
  \ebrule{
    \hypo{\Gamma \vdashT f : \Pi(A, B)}
    \infer1[($\eta$-rule)]{\Gamma \vdashT f = \lambda(A,B,\app(A\p,B\p^+,f\p, v_0)) : \Pi(A, B)}
  }
  \and
  \ebrule{
    \hypo{\Gamma \vdashT a : \U_n}
    \hypo{\Gamma.T_n(a) \vdashT b : \U_n}
    \infer2{\Gamma \vdashT T_n(\Pi^n(a, b)) = \Pi(T_n(a),T_n(b))}
  }
  \and
  \ebrule{
    \hypo{\Gamma \vdashT a : \U_n}
    \hypo{\Gamma.T_n(a) \vdashT b : \U_n}
    \infer2[($n\leq p$)]{\Gamma \vdashT t_n^p(\Pi^n(a, b)) = \Pi^p(t_n^p(a),t_n^p(b)) : \U_p}
  }
  \and
  \ebrule{
    \hypo{\Gamma \vdashT}
    \infer1[($m<n$)]{\Gamma \vdashT T_n(\U_m^n) = \U_m}
  }
  \and
  \ebrule{
    \hypo{\Gamma \vdashT}
    \infer1[($m<n\leq p$)]{\Gamma \vdashT t_n^p(\U_m^n) = \U_m^p : \U_p}
  }
  \and
  \ebrule{
    \hypo{\Gamma \vdashT a : \U_m}
    \infer1[($m \leq n$)]{\Gamma \vdashT T_n(t_m^n(a)) = T_m(a)}
  }
  \and
  \ebrule{
    \hypo{\Gamma \vdashT a : \U_m}
    \infer1[($m\leq n\leq p$)]{\Gamma \vdashT t_n^p(t_m^n(a)) = t_m^p(a) : \U_p}
  }
  \and
  \ebrule{
    \hypo{\Gamma \vdashT a : \U_n}
    \infer1{\Gamma \vdashT t_n^n(a) = a : \U_n}
  }
\end{mathparsmall}

\subsection{The system \ThyRussell : Russell with annotations}

\begin{mathparsmall}
  \ebrule{
    \hypo{\Gamma \vdashR A \;\type}
    \infer1{\Gamma.A \vdashR v_0 : A\p}
  }
  \and
  \ebrule{
    \hypo{\Gamma \vdashR v_n : A}
    \hypo{\Gamma \vdashR B \;\type}
    \infer2{\Gamma.B \vdashR v_{n+1} : A\p}
  }
  \and
  \ebrule{
    \hypo{\Gamma \vdashR A \;\type}
    \hypo{\Gamma.A \vdashR B \;\type}
    \infer2{\Gamma \vdashR \Pi(A, B) \;\type}
  }
  \and
  \ebrule{
    \hypo{\Gamma.A \vdashR b : B}
    \hypo{\Gamma.A \vdashR B \;\type}
    \hypo{\Gamma \vdashR A \;\type}
    \infer3{\Gamma \vdashR \lambda(A,B,b) : \Pi(A, B)}
  }
  \and
  \ebrule{
    \hypo{\Gamma \vdashR c : \Pi(A, B)}
    \hypo{\Gamma \vdashR a : A}
    \hypo{\Gamma.A \vdashR B \;\type}
    \hypo{\Gamma \vdashR A \;\type}
    \infer4{\Gamma \vdashR \app(A,B,c,a) : B[a]}
  }
  \and
  \ebrule{
    \hypo{\Gamma \vdashR}
    \infer1{\Gamma \vdashR \U_n \;\type}
  }
  \and
  \ebrule{
    \hypo{\Gamma \vdashR A : \U_n}
    \infer1{\Gamma \vdashR A \;\type}
  }
  \and
  \ebrule{
    \hypo{\Gamma \vdashR A : \U_n}
    \hypo{\Gamma.A \vdashR B : \U_n}
    \infer2{\Gamma \vdashR \Pi(A, B) : \U_n}
  }
  \and
  \ebrule{
    \hypo{\Gamma \vdashR}
    \infer1[($m<n$)]{\Gamma \vdashR \U_m : \U_n}
  }
  \and
  \ebrule{
    \hypo{\Gamma \vdashR A : \U_n}
    \infer1[($n<p$)]{\Gamma \vdashR A : \U_p}
  }
  \and
  \ebrule{
    \hypo{\Gamma \vdashR a : A}
    \hypo{\Gamma \vdashR A = B}
    \infer2{\Gamma \vdashR a : B}
  }
  \and
  \ebrule{
    \hypo{\Gamma \vdashR a = a' : A}
    \hypo{\Gamma \vdashR A = B}
    \infer2{\Gamma \vdashR a = a' : B}
  }
  \and
  \ebrule{
    \hypo{\Gamma \vdashR A \;\type}
    \infer1{\Gamma \vdashR A = A}
  }
  \and
  \ebrule{
    \hypo{\Gamma \vdashR A = B}
    \hypo{\Gamma \vdashR A = C}
    \infer2{\Gamma \vdashR B = C}
  }
  \and
  \ebrule{
    \hypo{\Gamma \vdashR a : A}
    \infer1{\Gamma \vdashR a = a : A}
  }
  \and
  \ebrule{
    \hypo{\Gamma \vdashR a = b : A}
    \hypo{\Gamma \vdashR a = c : A}
    \infer2{\Gamma \vdashR b = c : A}
  }
  \and
  \ebrule{
    \hypo{\Gamma \vdashR A = A'}
    \hypo{\Gamma.A \vdashR B = B'}
    \infer2{\Gamma \vdashR \Pi(A, B) = \Pi(A', B')}
  }
  \and
  \ebrule{
    \hypo{\Gamma \vdashR A = A' : \U_n}
    \hypo{\Gamma.A \vdashR B = B' : \U_n}
    \infer2{\Gamma \vdashR \Pi(A, B) = \Pi(A', B') : \U_n}
  }
  \and
  \ebrule{
    \hypo{\Gamma \vdashR A = A'}
    \hypo{\Gamma.A \vdashR B = B'}
    \hypo{\Gamma.A \vdashR b = b' : B}
    \infer3{\Gamma \vdashR \lambda(A,B,b) = \lambda(A',B',b') : \Pi(A, B)}
  }
  \and
  \ebrule{
    \hypo{\Gamma \vdashR A = A'}
    \hypo{\Gamma.A \vdashR B = B'}
    \hypo{\Gamma \vdashR c = c' : \Pi(A, B)}
    \hypo{\Gamma \vdashR a = a' : A}
    \infer4{\Gamma \vdashR \app(A,B,c,a) = \app(A',B',c',a') : B[a]}
  }
  \and
  \ebrule{
    \hypo{\Gamma.A \vdashR b : B}
    \hypo{\Gamma \vdashR a : A}
    \infer2[($\beta$-rule)]{\Gamma \vdashR \app(A,B,\lambda(A,B,b),a) = b[a] : B[a]}
  }
  \and
  \ebrule{
    \hypo{\Gamma \vdashR f : \Pi(A, B)}
    \infer1[($\eta$-rule)]{\Gamma \vdashR f = \lambda(A,B,\app(A\p,B\p^+,f\p, v_0)) : \Pi(A, B)}
  }
  \and
  \ebrule{
    \hypo{\Gamma \vdashR A = A' : \U_n}
    \infer1{\Gamma \vdashR A = A'}
  }
  \and
  \ebrule{
    \hypo{\Gamma \vdashR A = A' : \U_n}
    \infer1[($n<p$)]{\Gamma \vdashR A = A' : \U_p}
  }
\end{mathparsmall}

\subsection{The system \ThyPTS : Russell without annotations}

\begin{mathparsmall}
  \ebrule{
    \hypo{\Gamma \vdashP A \;\type}
    \infer1{\Gamma.A \vdashP v_0 : A\p}
  }
  \and
  \ebrule{
    \hypo{\Gamma \vdashP v_n : A}
    \hypo{\Gamma \vdashP B \;\type}
    \infer2{\Gamma.B \vdashP v_{n+1} : A\p}
  }
  \and
  \ebrule{
    \hypo{\Gamma \vdashP A \;\type}
    \hypo{\Gamma.A \vdashP B \;\type}
    \infer2{\Gamma \vdashP \Pi(A, B) \;\type}
  }
  \and
  \ebrule{
    \hypo{\Gamma.A \vdashP b : B}
    \hypo{\Gamma.A \vdashP B \;\type}
    \hypo{\Gamma \vdashP A \;\type}
    \infer3{\Gamma \vdashP \lambda(A, b) : \Pi(A, B)}
  }
  \and
  \ebrule{
    \hypo{\Gamma \vdashP c : \Pi(A, B)}
    \hypo{\Gamma \vdashP a : A}
    \infer2{\Gamma \vdashP c\; a : B[a]}
  }
  \and
  \ebrule{
    \hypo{\Gamma \vdashP}
    \infer1{\Gamma \vdashP \U_n \;\type}
  }
  \and
  \ebrule{
    \hypo{\Gamma \vdashP A : \U_n}
    \infer1{\Gamma \vdashP A \;\type}
  }
  \and
  \ebrule{
    \hypo{\Gamma \vdashP A : \U_n}
    \hypo{\Gamma.A \vdashP B : \U_n}
    \infer2{\Gamma \vdashP \Pi(A, B) : \U_n}
  }
  \and
  \ebrule{
    \hypo{\Gamma \vdashP}
    \infer1[($m<n$)]{\Gamma \vdashP \U_m : \U_n}
  }
  \and
  \ebrule{
    \hypo{\Gamma \vdashP A : \U_n}
    \infer1[($n<p$)]{\Gamma \vdashP A : \U_p}
  }
  \and
  \ebrule{
    \hypo{\Gamma \vdashP a : A}
    \hypo{\Gamma \vdashP A = B}
    \infer2{\Gamma \vdashP a : B}
  }
  \and
  \ebrule{
    \hypo{\Gamma \vdashP a = a' : A}
    \hypo{\Gamma \vdashP A = B}
    \infer2{\Gamma \vdashP a = a' : B}
  }
  \and
  \ebrule{
    \hypo{\Gamma \vdashP A \;\type}
    \infer1{\Gamma \vdashP A = A}
  }
  \and
  \ebrule{
    \hypo{\Gamma \vdashP A = B}
    \hypo{\Gamma \vdashP A = C}
    \infer2{\Gamma \vdashP B = C}
  }
  \and
  \ebrule{
    \hypo{\Gamma \vdashP a : A}
    \infer1{\Gamma \vdashP a = a : A}
  }
  \and
  \ebrule{
    \hypo{\Gamma \vdashP a = b : A}
    \hypo{\Gamma \vdashP a = c : A}
    \infer2{\Gamma \vdashP b = c : A}
  }
  \and
  \ebrule{
    \hypo{\Gamma \vdashP A = A'}
    \hypo{\Gamma.A \vdashP B = B'}
    \infer2{\Gamma \vdashP \Pi(A, B) = \Pi(A', B')}
  }
  \and
  \ebrule{
    \hypo{\Gamma \vdashP A = A' : \U_n}
    \hypo{\Gamma.A \vdashP B = B' : \U_n}
    \infer2{\Gamma \vdashP \Pi(A, B) = \Pi(A', B') : \U_n}
  }
  \and
  \ebrule{
    \hypo{\Gamma \vdashP A = A'}
    \hypo{\Gamma.A \vdashP b = b' : B}
    \infer2{\Gamma \vdashP \lambda(A, b) = \lambda(A', b') : \Pi(A, B)}
  }
  \and
  \ebrule{
    \hypo{\Gamma \vdashP c = c' : \Pi(A, B)}
    \hypo{\Gamma \vdashP a = a' : A}
    \infer2{\Gamma \vdashP c\; a = c'\; a' : B[a]}
  }
  \and
  \ebrule{
    \hypo{\Gamma.A \vdashP b : B}
    \hypo{\Gamma \vdashP a : A}
    \infer2[($\beta$-rule)]{\Gamma \vdashP (\lambda(A, b)) \; a = b[a] : B[a]}
  }
  \and
  \ebrule{
    \hypo{\Gamma \vdashP f : \Pi(A, B)}
    \infer1[($\eta$-rule)]{\Gamma \vdashP f = \lambda(A, f\p \; v_0) : \Pi(A, B)}
  }
  \and
  \ebrule{
    \hypo{\Gamma \vdashP A = A' : \U_n}
    \infer1{\Gamma \vdashP A = A'}
  }
  \and
  \ebrule{
    \hypo{\Gamma \vdashP A = A' : \U_n}
    \infer1[($n<p$)]{\Gamma \vdashP A = A' : \U_p}
  }
\end{mathparsmall}

\end{document}